\documentclass[12pt]{article}
\usepackage{amsmath,amsfonts,amsthm,amssymb}
\usepackage[doublespacing]{setspace}
\usepackage[hmargin={1.0in, 1.0in}, vmargin={1.0in, 1.0in}]{geometry}
\usepackage[utf8x]{inputenc}
\usepackage[english]{babel}
\usepackage{fancyhdr}
\usepackage{lastpage}
\usepackage{extramarks}
\usepackage{chngpage}
\usepackage{soul,color}
\usepackage{graphicx,float,wrapfig}
\usepackage{layout}
\usepackage{fullpage}
\usepackage{natbib, bm}
\usepackage{lscape}
\usepackage{tikz}
\usepackage{titlesec}
\usepackage{subcaption}
\usepackage{tabu}

\titlespacing\section{0pt}{2pt plus 2pt minus 2pt}{0pt plus 1pt minus 2pt}
\titlespacing\subsection{0pt}{1pt plus 0pt minus 1pt}{0pt plus 0pt minus 1pt}
\titlespacing\subsubsection{0pt}{1pt plus 0pt minus 1pt}{0pt plus 0pt minus 1pt}
\def\bfy{\mathbf y}

\def\bfF{\mathbf F}

\def\bfZ{\mathbf Z}
\def\bfY{\mathbf Y}

\def\bfbeta{\boldsymbol{\beta}}
\def\bfphi{\boldsymbol{\phi}}

\def\bfeta{\boldsymbol{\eta}}

\newtheorem{theorem}{Theorem}

\newtheorem{corollary}{Corollary}

\newtheorem{proposition}{Proposition}

\newtheorem*{definition*}{Definition}

\newcommand{\blind}{1}

\begin{document}
	\setlength{\abovedisplayskip}{3pt}
	\setlength{\belowdisplayskip}{3pt}
	\setlength{\abovedisplayshortskip}{2pt}
	\setlength{\belowdisplayshortskip}{2pt}	

\if1\blind
{
	\title{Modeling Multivariate Time Series with Copula-linked Univariate D-vines}	
	\author{
	Zifeng Zhao\\
	Mendoza College of Business, University of Notre Dame
    \and
	Peng Shi\\
	Wisconsin School of Business, University of Wisconsin-Madison
    \and
	Zhengjun Zhang\\
    Department of Statistics, University of Wisconsin-Madison}
	\date{}	
	\maketitle
} \fi

\if0\blind
{
	\title{Modeling Multivariate Time Series with Copula-linked Univariate D-vines}
	\author{}
	\date{}
	\maketitle
} \fi

\begin{abstract}
	This paper proposes a novel multivariate time series model named Copula-linked univariate D-vines~(CuDvine), which enables the simultaneous copula-based modeling of both temporal and cross-sectional dependence for multivariate time series. To construct CuDvine, we first build a semiparametric univariate D-vine time series model~(uDvine) based on a D-vine. The uDvine generalizes the existing first-order copula-based Markov chain models to Markov chains of an arbitrary-order. Building upon uDvine, we construct CuDvine by linking multiple uDvines via a parametric copula. As a simple and tractable model, CuDvine provides flexible models for marginal behavior and temporal dependence of time series, and can also incorporate sophisticated cross-sectional dependence such as time-varying and spatio-temporal dependence for high-dimensional applications. Robust and computationally efficient procedures, including a sequential model selection method and a two-stage MLE, are proposed for model estimation and inference, and their statistical properties are investigated. Numerical experiments are conducted to demonstrate the flexibility of CuDvine, and to examine the performance of the sequential model selection procedure and the two-stage MLE. Real data applications on the Australian electricity price data demonstrate the superior performance of CuDvine to traditional multivariate time series models.
\end{abstract}

\noindent\textit{Keywords}: multivariate time series, D-vine, time-varying dependence, spatio-temporal dependence, Markov chains, two-stage maximum likelihood estimation

\section{Introduction}
Modeling dependence for multivariate time series is essential to statistical applications in various fields. For instance, see \cite{Patton2012}, \cite{Brechmann2012}, \cite{Nikoloulopoulos2012} and \cite{Zhao2020} in finance, \cite{Smith2015} and \cite{Smith2016} in economics, and \cite{Erhardt2015} in climate monitoring. Roughly speaking, there are two types of dependence embedded in multivariate time series. One is the temporal dependence within each component univariate time series. The other is the cross-sectional dependence across all the component univariate time series. Multivariate time series often presents complicated dependence structures, such as nonlinear dependence, tail dependence, as well as asymmetric dependence, which makes dependence modeling a challenging yet crucial task. A desirable feature of a multivariate time series model is being able to accommodate the complex dependence in both temporal and cross-sectional dimension.

In the literature, copula is one of the most widely used tools for introducing flexible dependence structures among multivariate outcomes. A $d$-dimensional copula is a multivariate distribution function on $(0, 1)^d$ with uniform margins. By \cite{Sklar1959}'s theorem, any multivariate distribution $\bfF$ can be separated into its marginals $(F_1,\ldots, F_d)$ and a copula $C$, where the copula captures all the scale-free dependence of the multivariate distribution. In particular, suppose there is a random vector $\bfY \in \mathbb{R}^d$ such that $\bfY$ follows $\bfF$, we have $\bfF(\bfy)=C(F_1(y_1), \ldots, F_d(y_d))$, where $\bfy=(y_1,\ldots, y_d)'$ is a realization of $\bfY$. If all the marginals of $\bfF$ are absolutely continuous, the copula $C$ is unique.

Most existing copula-based time series models focus on the cross-sectional dependence of multivariate time series, see, for example, the semiparametric copula-based multivariate dynamic models~(SCOMDY) in \cite{ChenFan2006a}. Under the SCOMDY framework, standard univariate time series models, such as ARMA and GARCH~\citep{Engle1982, Bollerslev1986}, are used to capture the temporal dependence in the conditional mean and variance of each component univariate time series. A parametric copula is then used to specify the cross-sectional dependence across the {standardized} innovations of all the component univariate time series. See \cite{Patton2006a}, \cite{Brechmann2012}, \cite{Almeida2012} and \cite{OhPatton2015} for related models under the SCOMDY framework. \textcolor{black}{\cite{Oh2018} further extends the SCOMDY framework by allowing a high-dimensional time-varying cross-sectional copula.}

Using copulas to model the temporal dependence of univariate time series is not uncommon. \cite{ChenFan2006b} and \cite{Domma2009} consider copula-based Markov chains, where copulas and flexible marginal distributions are used to specify the transitional probability of the Markov chains. \cite{Ibragimov2009}, \cite{Chen2009} and \cite{Beare2010} study the probabilistic properties of copula-based Markov chains. \cite{Birr2017} propose a copula spectral method for studying variation in temporal dependence structure. See \cite{Joe2014} for a nice presentation of copula-based Markov chains. However, most of the literature focus on first-order Markov chains using bivariate copulas, possibly due to the variety of choices and mathematical tractability in the low dimensional setting.

To extend the copula-based univariate time series model to higher-order Markov chains, a framework to generate flexible yet tractable multivariate copulas is required. A promising direction is vine-copula{~\citep[see][]{Joe1996,Bedford2002,Aas2009}, which generates multivariate copulas based on iterative pairwise construction of bivariate copulas. See \cite{Kurowicka2006} and \cite{Kurowicka2011} for more details of vine-copula.} The D-vine, a specially structured vine-copula, is of particular interest due to its simplicity and natural interpretation under time series setting. \cite{Smith2010} and \cite{Shi2017} employ D-vine to account for the temporal dependence in longitudinal data, and \cite{LoaizaMaya2017} use D-vine to capture the temporal dependence in stationary heteroskedastic time series. \textcolor{black}{A brief technical review of D-vine is given in Section 2.1. }

Although copulas have been proposed for modeling temporal and cross-sectional dependence in the aforementioned two separate strands of studies, there are few multivariate time series models that use copulas to account for both types of dependence simultaneously. Some notable exceptions are: \cite{Smith2015} and \cite{Beare2015} first stack the multivariate time series into a univariate time series and then design D-vine based dependence structures for the resulted univariate time series; \cite{Brechmann2014} use an R-vine to simultaneously model the temporal and cross-sectional dependence. These approaches demonstrate flexible dependence structures and show superior performance to the standard multivariate time series models, such as Vector AR, in various applications. One potential drawback is that these models are technically complicated and can be difficult to implement. For example, all the proposed methods involve a direct copula-based joint distribution of a high-dimensional vector of length $T\times d$, which is challenging both analytically and {computationally}, especially when the cross-sectional dimension $d$ is high. Another potential disadvantage is that it can be difficult for these models to impose parsimonious and intuitively interpretable structures into the cross-sectional dependence, such as time-varying, and spatial or factor-structured dependence, which may further hinder their abilities in modeling high-dimensional time series such as large panel data or spatio-temporal data.

In this paper, we aim to design a simple, intuitive and flexible multivariate time series model that enables the simultaneous copula-based modeling of both temporal and cross-sectional dependence, \textcolor{black}{and accommodates multivariate time series modeling in the high-dimensional setting.} Specifically, based on pair copula construction, we first design a semiparametric univariate D-vine time series model~(uDvine) that generalizes the existing first-order copula-based Markov chain to an arbitrary-order Markov chain. We then further propose a multivariate time series model named Copula-linked univariate D-vines~(CuDvine), where a parametric copula is employed to link multiple uDvines and specify the (conditional) cross-sectional dependence. \textcolor{black}{Flexible specification of this parametric copula is designed for modeling complex cross-sectional dependence, such as high-dimensional, time-varying or spatial dependence.} {Compared to existing copula-based multivariate time series models, a distinctive advantage of CuDvine is its flexibility in the specification of both copula-based temporal dependence and copula-based cross-sectional dependence. Because of this property, CuDvine extends the applicability of vine-copula based time series models to the important area of high-dimensional and spatio-temporal time series modeling. See more detailed comparisons between CuDvine and existing literature on copula-based multivariate time series modeling in Section 2.3.2.}

The main contributions of this paper are two-fold. In terms of statistical modeling, thanks to the use of a novel hybrid modeling approach, the proposed CuDvine achieves a nice balance between model flexibility and (analytical and computational) tractability. As demonstrated in real data applications, CuDvine can readily handle complicated marginal behavior and temporal dependence of time series, as well as model sophisticated \textcolor{black}{high-dimensional} cross-sectional dependence structures such as time-varying and parsimonious spatio-temporal dependence. In terms of statistical theory, we give a complete treatment of model selection and estimation for both uDvine and CuDvine, where robust and computationally efficient procedures are proposed. Although the idea of using D-vine to capture temporal dependence is not new, to our best knowledge, we are the first one to systematically study the probabilistic properties of D-vine based time series and the statistical properties of its estimators.

The rest of the paper is organized as follows. Section 2 presents uDvine and CuDvine, and investigates their probabilistic properties. In Section 3, a sequential model selection procedure and a two-stage maximum likelihood estimator~(MLE) are proposed for model inference and estimation. Their statistical properties are investigated as well. Numerical experiments are conducted in Section 4 to demonstrate the flexibility of CuDvine, and to examine the performance of the sequential model selection procedure and the two-stage MLE. Real data applications on the Australian electricity price are considered in Section 5, where significant improvement over traditional time series models is observed. We conclude the paper in Section 6. The supplementary material contains additional real data analysis, the proofs of the theorems and other technical materials.

\section{The D-vine based Time Series Models}
\subsection{Background}
\textcolor{black}{In this section, we give a brief technical review of D-vine, which serves as the building block of the later proposed uDvine and CuDvine.} According to \cite{Aas2009}, the density of a $T$-dimensional random vector $\bfY=\{Y_t\}_{t=1}^T \in \mathbb{R}^T$~{(here $\bfY$ denotes a univariate time series of length $T$)} based on D-vine is given by the $T$ marginal distributions $\{F_t(\cdot)\}_{t=1}^T$ of $\bfY$ and $T(T-1)/2$ bivariate copulas $\{\{ c_{s,t} \}_{s=1}^{t-1}\}_{t={2}}^T$ such that
\begin{align}
&f_D(\bfy;\bfbeta) = {f(y_1)}\prod_{t={2}}^{T}f(y_t|y_{t-1},\ldots, y_1)\nonumber\\
=&\prod_{t=1}^{T}f_t(y_t) {\prod_{t=2}^T}\prod_{s=1}^{t-1}c_{s,t}
(F_{s|(s+1):(t-1)}(y_s|y_{s+1}, \ldots, y_{t-1}), F_{t|(s+1):(t-1)}(y_t|y_{s+1}, \ldots, y_{t-1});\beta_{s,t}),
\label{Dvinepdf}
\end{align}
where $f_t(\cdot)$ is the pdf of $F_t(\cdot)$, $F_{s|(s+1):(t-1)}(y_s|y_{s+1}, \ldots, y_{t-1})$ and $F_{t|(s+1):(t-1)}(y_t|y_{s+1}, \ldots, y_{t-1})$ are conditional cdf of $Y_s$ and $Y_t$ given variables $(Y_{s+1},\ldots, Y_{t-1})$, and can be calculated recursively based on $\{F_t(\cdot)\}$ and $\{c_{s,t}\}$ by the algorithm in \cite{Aas2009}. \textcolor{black}{Here and after, we use the convention that $(s+1):(t-1)=\varnothing$ and $y_{s+1}, \ldots, y_{t-1}=\varnothing$ if $s+1>t-1.$} The parameter of the bivariate copula $c_{s,t}$ is denoted by $\beta_{s,t}$ and $\bfbeta=\{\{\beta_{s,t} \}_{s=1}^{t-1}\}_{t={2}}^T$.

\begin{figure}[h]
	\centering
	\includegraphics[scale=0.39]{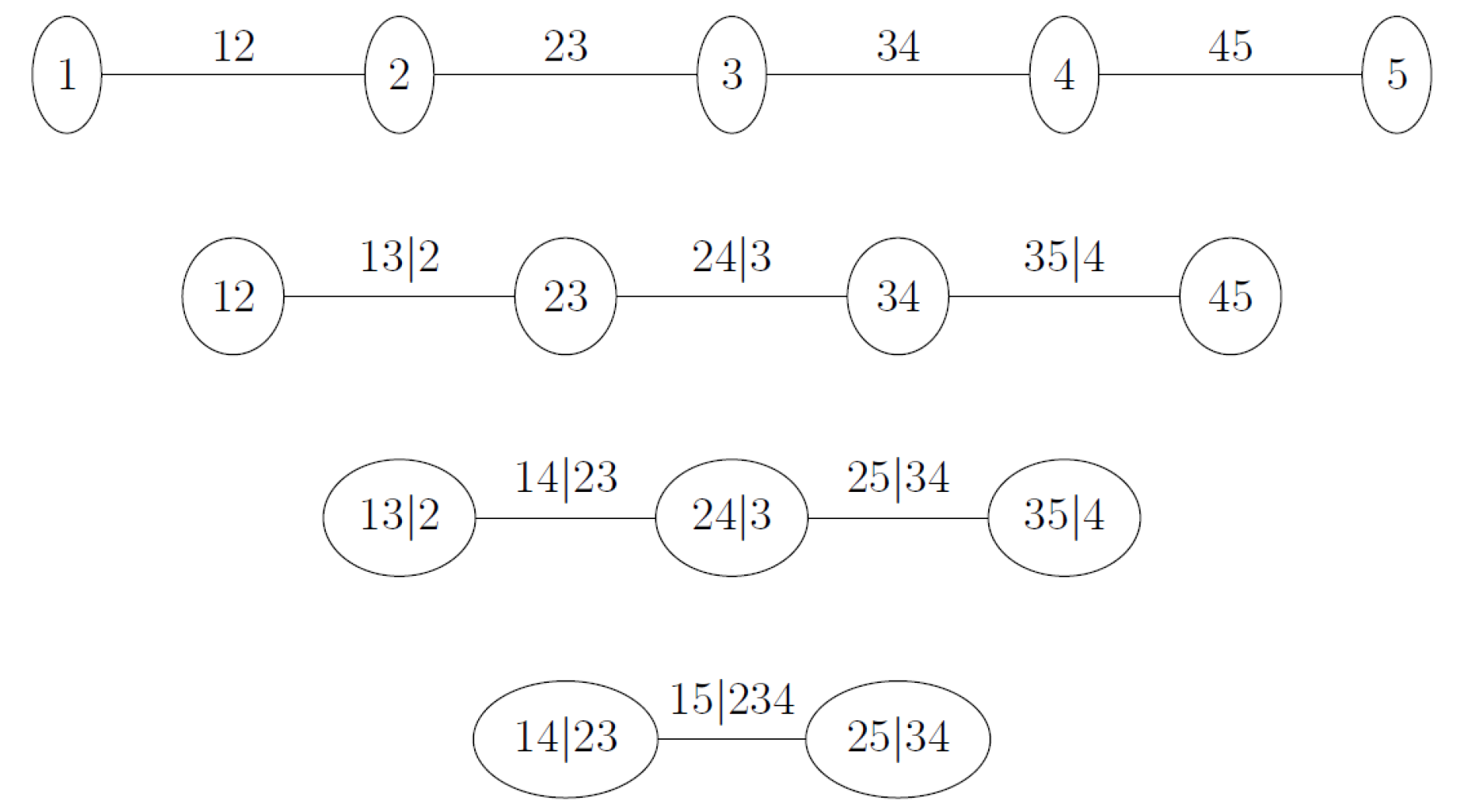}
	\caption{{\it A 5-dimension D-vine.}}
	\label{Dvine}
\end{figure}

An example of D-vine for $T=5$ is exhibited in Figure \ref{Dvine}. The nodes in tree 1~(top) represent the {probability integral transformed} marginals $\{F(Y_t)\}_{t=1}^T$ and the edges in each tree becomes the nodes in the next tree. From left to right, the $s$th edge in tree $t-s~(t>s)$ corresponds to the (conditional) bivariate copula $c_{s, t}$ that is used in $f_D(\bfy, \bfbeta)$ to specify the conditional joint distribution of $(Y_s, Y_t)$ given variables $(Y_{s+1},\ldots, Y_{t-1})$. The edges of the entire D-vine indicate the bivariate copulas $\{\{c_{s,t}\}_{s=1}^{t-1}\}_{t={2}}^T$ that contribute to the pair copula constructions. The key feature of D-vine is that the edges of each tree only connect adjacent nodes, which makes it simple to understand and naturally interpretable for time series. If $\bfY$ represents a univariate time series, D-vine provides a valid univariate time series model.

\subsection{Univariate D-vine time series model~(uDvine)}
In this section, we introduce the univariate D-vine time series model~(uDvine) and study its probabilistic properties. Throughout the section, we use $\bfY=\{Y_t\}_{t=1}^T$ to denote a univariate time series and we assume the time series is strictly stationary. Note that the general formula for the density of $\bfY=\{Y_t\}_{t=1}^T$ based on D-vine is given by $(\ref{Dvinepdf})$, which depends on $T$ marginal distributions $\{F_t(\cdot)\}_{t=1}^T$ of $\bfY$ and $T(T-1)/2$ bivariate copulas $\{\{ c_{s,t} \}_{s=1}^{t-1}\}_{t={2}}^T$.

\subsubsection{Model specification of uDvine}
The strict stationarity of $\{Y_t\}_{t=1}^T$ implies that the marginal distribution $F_t(\cdot)=F(\cdot)$ for all $t$ and that all bivariate copulas in the same tree must be identical, i.e. $c_{s,t}=c_{s',t'}$ if $t-s=t'-s'$. We call this the homogeneity condition. Thus, under the stationarity assumption, to fully specify the joint distribution of $\bfY$, one needs to specify a marginal distribution $F(\cdot)$ and $T-1$ bivariate copulas for tree 1 to tree $T-1$, which is unrealistic when $T$ is large.

A natural solution is to `truncate' D-vine after a certain level~(say tree $p$) and set all bivariate copulas beyond tree $p$, i.e. $\{c_{s, t}, t-s>p\}$, to be independent copulas\footnote{See \cite{Brechmann2012} and \cite{Brechmann2014} for a similar idea on truncating R-vine.}, where $p \ll T$. We call the univariate D-vine time series model truncated at tree $p$ the uDvine($p$) model. As shown later in Proposition \ref{pMC}, uDvine($p$) is a $p$-order homogeneous Markov chain. To maximize the flexibility of marginal behavior, we do not impose any parametric assumption on $F(\cdot)$ and only assume it to be absolutely continuous, which makes uDvine a semiparametric time series model~{(see Remark 1 for other choices of marginal distributions)}.

The joint distribution of $\{Y_t\}_{t=1}^T$ based on uDvine($p$) can be written as
\begin{align*}
&f_D(\bfy;\bfbeta)= {f(y_1)}\prod_{t={2}}^{T}f(y_t|y_{t-1},\ldots, y_1)={f(y_1)}\prod_{t={2}}^{T}f(y_t|y_{t-1},\ldots, y_{1 \vee (t-p)})\\ &=\prod_{t=1}^{T}f(y_t){\prod_{t=2}^T}\prod_{s=1\vee(t-p)}^{t-1}c_{s,t}(F_{s|(s+1):(t-1)}(y_s|y_{s+1},\cdots,y_{t-1}),F_{t|(s+1):(t-1)}(y_t|y_{s+1},\cdots,y_{t-1}); \beta_{s,t}),
\end{align*}
where $c_{s,t}$ is the bivariate copula in tree $t-s$ with parameter $\beta_{s,t}$, $F_{s|(s+1):(t-1)}$ and $F_{t|(s+1):(t-1)}$ are the conditional cdfs of $Y_s$ and $Y_t$ given $(Y_{s+1},\cdots, Y_{t-1})$. By the homogeneity condition, we have $F_{s|(s+1):(t-1)}=F_{s'|(s'+1):(t'-1)}$ and $F_{t|(s+1):(t-1)}=F_{t'|(s'+1):(t'-1)}$ for all $(s,t,s',t')$ such that $t'-s'=t-s$. We denote $\bfbeta=\{\beta_{s,t}\}$ as the collection of all parameters for the $p$ bivariate copulas and denote $\mathcal{F}_{t-1}=\sigma(Y_{t-1},Y_{t-2},\ldots)$.

For the purposes of estimation and prediction, the conditional distribution of uDvine is needed and can be easily derived from the joint distribution. By the Markovian property of uDvine($p$), it can be shown that, for $t>p$, the conditional pdf of $Y_t$ takes the form
\begin{align*}
&f(y_t|\mathcal{F}_{t-1})=f(y_t|y_{t-1},y_{t-2},\cdots,y_{t-p})\\
&=f(y_t)\cdot\prod_{s=t-p}^{t-1}c_{s,t}(F_{s|(s+1):(t-1)}(y_s|y_{s+1},\cdots,y_{t-1}),F_{t|(s+1):(t-1)}(y_t|y_{s+1},\cdots,y_{t-1}); \beta_{s,t}),
\end{align*}
which can be shown to be a function of $f(y_t)$, $\{F(y_{t-k})\}_{k=0}^p$ and $\bfbeta$. For simplicity of notation, we denote
\begin{align}
&w(F(y_{t}), F(y_{t-1}),\cdots, F(y_{t-p});\bfbeta)\nonumber\\
&=\prod_{s=t-p}^{t-1}c_{s,t}(F_{s|(s+1):(t-1)}(y_s|y_{s+1},\cdots,y_{t-1}),F_{t|(s+1):(t-1)}(y_t|y_{s+1},\cdots,y_{t-1}); \beta_{s,t}),
\label{Dvinepdf2}
\end{align}
where $w(u_1,u_{2},\cdots,u_{p+1};\bfbeta)$ can be derived\footnote{\label{uDvine2Demo}See Section \S2 of the supplementary material for the derived formulas for a uDvine(2).} based on the algorithm in \cite{Aas2009}. Together, we have $f(y_t|y_{t-1},y_{t-2},\cdots,y_{t-p})=f(y_t)\cdot w(F(y_{t}), F(y_{t-1}),\cdots, F(y_{t-p}); \bfbeta)$.

Similarly, it can be shown that, for $t>p$, the conditional cdf of $Y_t$ given $\mathcal{F}_{t-1}$ is a function of $\{F(y_{t-k})\}_{k=0}^p$ and $\bfbeta$. To simplify notation, we denote
\begin{align}
\label{gfunction}
{F}(y_{t}|\mathcal{F}_{t-1})=F(y_{t}|y_{t-1},\cdots, y_{t-p})=g(F(y_{t}), F(y_{t-1}),\cdots, F(y_{t-p}); \bfbeta),
\end{align}
where $g(u_1,\cdots,u_{p+1};\bfbeta)$ can also be derived\textsuperscript{\ref{uDvine2Demo}} based on the algorithm in \cite{Aas2009}.

Unlike many ``conditional'' univariate time series models, such as ARMA and GARCH, uDvine directly specifies the joint distribution of the univariate time series, instead of specifying the conditional distribution of $Y_t$ given $\mathcal{F}_{t-1}$. Most univariate time series models that are based on the conditional approach specify the temporal dependence via first and second order moments, which can be restrictive. On the contrary, uDvine does not impose constraints on either the marginal behavior of $Y_t$ or the temporal dependence due to the use of the semiparametric D-vine. Depending on the choices of bivariate copulas in each tree, uDvine can generate nonlinear, asymmetric, and tail dependence. The flexibility of uDvine is demonstrated through numerical experiments in Section 4.1 and through real data applications in Section 5.

The uDvine($p$) is a general model that nests many commonly used time series models as special cases. All the first-order copula-based Markov chains, e.g. \cite{ChenFan2006b}, are essentially a uDvine(1). In fact, all the stationary first-order Markov chains in $\mathbb{R}$, e.g. AR(1) models and ARCH(1) models in \cite{Engle1982}, are special cases of uDvine(1). Another important special case of uDvine($p$) is a stationary AR($p$) process with Gaussian innovations. {\cite{LoaizaMaya2017} show numerically that certain D-vine based time series model can generate volatility clustering effects as in GARCH model, Example 3 in Section \S1 of the supplementary material gives an analytical explanation of such phenomenon.}

{\textbf{Remark 1}: One advantage of copula-based modeling, and thus uDvine, is that it allows flexible specification of marginal distributions. In this paper, we use nonparametric marginal distributions to achieve maximum flexibility. To handle heavy-tailed time series in certain financial/economics applications, an alternative strategy is to employ a generalized Pareto distribution~(GPD) based semiparametric marginal distribution, see for example \cite{McNeil2000} for more details.}

\subsubsection{Stationarity and ergodicity of uDvine}
Note that under the homogeneity condition, the univariate time series $\{Y_t\}$ generated by uDvine($p$) is strictly stationary. In this section, we study the probabilistic properties of uDvine and show that under certain conditions, $\{Y_t\}$ is ergodic. {To our best knowledge, this is the first formal result on ergodicity of D-vine based time series, which extends the result of first-order copula-based Markov chains in \cite{ChenFan2006b}.}
\begin{proposition}
	Under the homogeneity condition, the univariate time series $\{Y_t\}$ generated by uDvine($p$) is a $p$-order homogeneous Markov chain.
	\label{pMC}
\end{proposition}

Proposition \ref{pMC} is in line with the Markov properties of D-vine studied in \cite{Smith2015} and \cite{Beare2015}. By Proposition \ref{pMC}, if we define $X_t=(F(Y_t), F(Y_{t-1}),\ldots, F(Y_{t-p+1}))$, the new process $\{X_t\}$ is a first-order homogeneous Markov chain with state space $(0,1)^p$. Since the marginal distribution $F(\cdot)$ of uDvine is absolutely continuous, we know that $F(Y_t)$ marginally follows the uniform distribution on $(0, 1)$. As noted in \cite{ChenFan2006b}, the stationarity and ergodicity of $\{Y_t\}$ and $\{F(Y_t)\}$ are equivalent due to the absolute continuity of the marginal distribution $F(\cdot)$. Theorem \ref{MCergodic} gives sufficient conditions for the ergodicity of $\{X_t\}$ and thus that of $\{Y_t\}$.
\begin{theorem}
	Under the homogeneity condition and Assumptions S.1 and S.2 in Section \S3 of the supplementary material, $\{X_t\}$ is positive Harris recurrent and geometrically ergodic, thus is $\{Y_t\}$, which follows uDvine($p$).
	\label{MCergodic}
\end{theorem}

A direct result of Theorem \ref{MCergodic} is the $\beta$-mixing property of uDvine($p$).
\begin{corollary}
	If Theorem \ref{MCergodic} holds, uDvine$(p)$ is $\beta$-mixing with an exponential decaying rate.
	\label{betaMixing}
\end{corollary}

\vspace{-1cm}

\subsection{Copula-linked univariate D-vines~(CuDvine) time series model}
The proposed uDvine accounts for various marginal behavior and temporal dependence of the univariate time series. To develop a flexible multivariate time series model, we employ an additional copula to specify the cross-sectional dependence across uDvines and propose the Copula-linked univariate D-vines~(CuDvine) time series model.

Throughout this section, $\{\bfY_t=(Y_{t1}, \ldots, Y_{td})\}_{t=1}^T$ denotes a $d$-dimensional multivariate time series, $\mathcal{F}_{t-1}=\sigma(\bfY_{t-1}, \bfY_{t-2},\ldots)$ denotes the sigma field of all past information and $\mathcal{F}_{t-1}^i=\sigma(Y_{t-1,i},Y_{t-2,i},\ldots)$ denotes the sigma field of the past information from the $i$th component univariate time series.

The time series $\{\bfY_t\}_{t=1}^T$ is defined as a CuDvine if its component univariate time series $\{Y_{ti}\}_{t=1}^T$ follows a uDvine($p_i$), for $i=1,\ldots, d$, and the conditional joint distribution $\bfF(\cdot|\mathcal{F}_{t-1})$ of $\bfY_t$ given $\mathcal{F}_{t-1}$ can be written as
\begin{align}
\label{CuDvineCDF}
\bfF(\bfy_{t}|\mathcal{F}_{t-1})= \bfF(y_{t1},\ldots,y_{td}|\mathcal{F}_{t-1})=C(F_1(y_{t1}|\mathcal{F}_{t-1}^1), \ldots, F_d(y_{td}|\mathcal{F}_{t-1}^d); \mathcal{F}_{t-1}, \gamma),
\end{align}
where $C(\cdot; \mathcal{F}_{t-1}, \gamma)$ is a $d$-dimensional {parametric} copula with parameter $\gamma$ that captures the conditional cross-sectional dependence given history $\mathcal{F}_{t-1}$, and $F_i(\cdot|\mathcal{F}_{t-1}^i)$ are the conditional marginal distribution of $Y_{ti}$ given its own history $\mathcal{F}_{t-1}^i$.

Since uDvine($p_i$) is a $p_i$-order Markov chain, we have $F_i(y_{ti}|\mathcal{F}_{t-1}^i)=F_i(y_{ti}|y_{t-1,i},\ldots, y_{t-{p_i},i})$. Given the marginal distribution $F_i(\cdot)$ and the parameter $\bfbeta_i$ of the bivariate copulas in the $i$th uDvine($p_i$), $F_i(y_{ti}|\mathcal{F}_{t-1}^i)$ is a function of $\{F_i(y_{t-k,i})\}_{k=0}^{p_i}$ and $\bfbeta_i$ such that
\begin{align}
	{F}_i(y_{ti}|\mathcal{F}_{t-1}^i)=F_i(y_{ti}|y_{t-1,i},\ldots, y_{t-{p_i},i})=g_i(F_i(y_{ti}), F_i(y_{t-1,i}),\ldots, F_i(y_{t-{p_i},i}); \bfbeta_i),
	\label{conditionalCDF}
\end{align}
where $g_i(u_1,\ldots,u_{{p_i}+1};\bfbeta_i)$ is defined in (\ref{gfunction}) in Section 2.2.1. In the following, without loss of generality, we assume that the order of all uDvines to be $p$.

{Note that $(\ref{CuDvineCDF})$ is a direct result of the conditional Sklar's theorem in \cite{Patton2006a}~(Theorem 1), which states that given \textit{any} $d$ conditional marginal distributions $F_i(\cdot|\mathcal{F}_{t-1}^i), i=1,\dots, d$ and \textit{any} conditional copula $C(\cdot; \mathcal{F}_{t-1}, \gamma)$, the function $\bfF(\bfy_{t}|\mathcal{F}_{t-1})$ in \eqref{CuDvineCDF} gives a valid $d$-dimensional conditional joint distribution of $\bfY_t$ given $\mathcal{F}_{t-1}$.}

{Importantly, this indicates that the parametric form of the conditional cross-sectional copula $C(\cdot; \mathcal{F}_{t-1}, \gamma)$ is not restricted and can be any copula, which greatly increase the flexibility of CuDvine. This is an important difference between CuDvine and existing vine-copula based multivariate time series models where both temporal and cross-sectional dependence are limited to D-vine copulas, see for example \cite{Beare2015} and \cite{Smith2015}.}

The specification of the cross-sectional copula $C(\cdot;\mathcal{F}_{t-1}, \gamma)$ is flexible and can take a variety of forms. A popular assumption in the multivariate time series literature is that the conditional copula of $\bfY_t$ given $\mathcal{F}_{t-1}$ does not depend on $\mathcal{F}_{t-1}$, which implies that $C(\cdot;\mathcal{F}_{t-1}, \gamma)$ is a static copula $C(\cdot; \gamma)$. For low-dimensional applications, $C(\cdot; \gamma)$ can be an unstructured copula such as elliptical copula or Archimedean copula. For high-dimensional applications, $C(\cdot; \gamma)$ can be a parsimonious factor-structured or spatial-structured copula. A time-varying $C(\cdot;\mathcal{F}_{t-1}, \gamma)$ where the cross-sectional dependence evolves according to $\mathcal{F}_{t-1}$ can also be readily implemented. \textcolor{black}{See Section 2.3.1 for more discussion of CuDvine for high-dimensional time series.} In real data analysis, we demonstrate the applications of CuDvine with both time-varying and spatial-structured cross-sectional copulas.

One implicit assumption of CuDvine is a conditional independence assumption --- (\textbf{A1}) the conditional marginal distribution of the $i$th component univariate time series $Y_{ti}$ given $\mathcal{F}_{t-1}$ only depends on its own history $\mathcal{F}_{t-1}^i$. \textbf{A1} may appear to be restrictive. However, plenty of multivariate time series models based on \textbf{A1} are shown to perform well in real data applications, see, for example, the SCOMDY framework in \cite{ChenFan2006a}, \cite{Patton2006a}, \cite{Dias2010}, \cite{Almeida2012}, and \cite{OhPatton2015}. See \cite{Shi2018} and \cite{Nikoloulopoulos2017} for models with \textbf{A1} for multivariate discrete or mixed longitudinal data. One advantage of \textbf{A1} is that it drastically reduces the number of parameters for temporal dependence from $O((dp)^2)$ to $O(dp)$ and enables the use of two-stage MLE. Together with the parsimonious structure of the cross-sectional copula, CuDvine can easily handle high-dimensional multivariate time series such as spatio-temporal data and large panel data of stock returns.

\subsubsection{\textcolor{black}{CuDvine for high-dimensional time series}}
\textcolor{black}{With the increasing availability of large financial datasets thanks to the advances of computing technologies, high-dimensional time series modeling has become an important topic~\citep{Fan2011}. In this section, we discuss two strategies for CuDvine to model high-dimensional time series. The essential idea is to use an elliptical copula, such as Gaussian or $t$-copula, for the conditional cross-sectional copula $C(\cdot;\mathcal{F}_{t-1},\gamma)$ and adapt the parametric specification of its correlation matrix $R$ to high-dimension.}

\textcolor{black}{The first strategy is via factor model, where we set $C(\cdot;\mathcal{F}_{t-1},\gamma)=C(\cdot;\gamma)$ to be a static elliptical copula and impose a factor structure on its correlation matrix $R.$ Factor model is arguably the most popular approach for handling high-dimensional time series, see \cite{Bai2002} and \cite{Lam2012}. Here, we adapt the block factor model proposed in \cite{OhPatton2015} and \cite{Zhao2017}, which is designed specifically for financial data. A $d$-dimensional random vector $\bfZ=(Z_1,Z_2,\cdots,Z_d)$ follows a block factor model if it can be grouped into $m$ blocks such that $\bfZ=\bigcup_{i=1}^m(Z_{i1},\cdots,Z_{id_i})$ with $\sum_{i=1}^m d_i=d$ and
\begin{align*}
	Z_{ij}=\phi_{i0} B_0+\phi_{i1} B_i+\epsilon_{ij},~i=1,\cdots,m,~j=1,\cdots, d_i,
\end{align*}
where $B_0$ is the common factor across all blocks, $B_i$s are block-specific factors, $\epsilon_{ij}$s are subject level noise, and all random variables are mutually independent with unit variance. The correlation matrix $R$ implied by $\bfZ$ admits a block factor structure with
\begin{align}\label{blockfactor}
	&Cor(Z_{ij}, Z_{ij'})=\frac{\phi_{i0}^2+\phi_{i1}^2}{1+\phi_{i0}^2+\phi_{i1}^2}, \text{~ for } i=1,\cdots,m, \text{ and } j\neq j', \nonumber\\
	&Cor(Z_{ij}, Z_{i'j'})=\frac{\phi_{i0}\phi_{i'0}}{\sqrt{1+\phi_{i0}^2+\phi_{i1}^2}\sqrt{1+\phi_{i'0}^2+\phi_{i'1}^2}}, \text{~ for } i=1,\cdots, m, \text{ and } i'\neq i.
\end{align}
As shown in \cite{OhPatton2015} and \cite{Zhao2017}, the block factor model is intuitive and is proven to be effective for modeling high-dimensional financial time series as assets~(e.g. stocks) can be naturally grouped based on industrial sectors. In Section 4.4, we further conduct numerical experiments to demonstrate the promising ability of block-factor structured CuDvine to model high-dimensional multivariate GARCH process.}

\textcolor{black}{The second strategy is via shrinkage, where we set $C(\cdot;\mathcal{F}_{t-1},\gamma)$ to be a time-varying elliptical copula and impose a shrinkage-DCC structure on its conditional correlation matrix $R_t.$ The shrinkage-DCC model is proposed in \cite{Engle2019} for modeling high-dimensional time series, where a shrinkage estimator is used to recover the unconditional correlation matrix of the original DCC model in \cite{Engle2002}, and is shown to work well for modeling high-dimensional asset returns. In Section 5, we illustrate the promising performance of a DCC-structured CuDvine~(adapting the original DCC model in \cite{Engle2002}) on modeling electricity prices of multiple regions in Australia.}

\subsubsection{{Relationship with existing modeling approaches}}
Most existing multivariate time series models, such as the SCOMDY framework, follow a purely ``conditional" modeling approach in the sense that both the temporal and cross-sectional dependence are specified via conditional distributions of $\bfY_t$ given $\mathcal{F}_{t-1}$. As discussed in Section 2.2.1 and noted by \cite{Smith2016}, the conditional approach can be restrictive in terms of modeling the marginal behavior and temporal dependence of the component univariate time series.

{In contrast, the copula time series models} in \cite{Brechmann2014}, \cite{Beare2015} and \cite{Smith2015} follow a purely ``joint" modeling approach in the sense that the joint distribution of all the $Td$ observations of $\{\bfY_t\}_{t=1}^T$ are specified directly, which helps offer great modeling flexibility. On the other hand, the joint approach is computationally and analytically complicated, and can be difficult to incorporate structured cross-sectional dependence such as time-varying and factor/spatial-structured dependence, {which limits its applicability to low-dimensional time series with a small cross-sectional dimension $d$.}

CuDvine follows a unique ``hybrid'' modeling approach -- the marginal behavior and temporal dependence are modeled by a joint approach via uDvine, and the cross-sectional dependence is modeled by a conditional approach via a $d$-dimensional copula. The D-vine based joint approach for the component univariate time series allows CuDvine to accommodate sophisticated marginal behavior and temporal dependence, which is demonstrated later by numerical experiments and real data applications. The copula-based conditional approach enables CuDvine to generate flexible cross-sectional dependence and makes the estimation and prediction procedure straightforward and computationally efficient, {which facilitates its application to high-dimensional time series. CuDvine can readily model time-varying cross-sectional dependence and high-dimensional factor/spatio-temporal dependence as demonstrated in Sections 4 and 5.} To summarize, the novel hybrid modeling approach makes CuDvine achieve highly flexible modeling ability and remain analytically and computationally tractable.

\section{Estimation and Inference}
As pointed out by \cite{Aas2009}, the inference for D-vine consists of two parts: (a) the choice of bivariate copula types and (b) estimation of copula parameters. The same tasks apply to CuDvine. In Section 3.1, we discuss model selection for CuDvine. In particular, we propose a sequential model selection procedure for the component uDvine. In Section 3.2, we propose a two-stage MLE for parameter estimation of a given CuDvine.

\subsection{Selection of bivariate copulas for uDvine}
To implement a CuDvine, one needs to specify the order $p$ and the bivariate copulas $\{c_{s,t}\}$ for each component uDvine, and one also needs to specify the cross-sectional copula $C(\cdot)$. The selection of $C(\cdot)$ can rely on standard procedures such as AIC or BIC. Here, we focus on the model selection for the component uDvine.

Given a set of candidate copulas~(say $m$ different copulas) and an order $p$, the number of possible uDvines is $m^p$, which can be quite large even for moderate $m$ and $p$. For computational feasibility, we propose a tree-by-tree sequential selection procedure.

The basic procedure is as follows. We start with the first tree, selecting the appropriate copula from a given set of candidates and estimating its parameters. Fixing the selected copula and its estimated parameters in the first tree, we then select the optimal copula and estimate its dependence parameters for the second tree. We continue this process for the next tree of a higher order while holding the selected copulas and the corresponding estimated parameters fixed in all previous trees. If an independent copula is selected for a certain tree, we then truncate the uDvine, i.e. assume conditional independence in all higher order trees~\citep[see, for example, ][]{Brechmann2012}.

The commonly used BIC is employed for the copula selection for each tree. As shown in Section 4.2, the sequential model selection procedure for uDvine is computationally efficient and can identify the true model accurately.

{\color{black} \textbf{Remark 2}: Given i.i.d. random vectors generated by a D-vine with known bivariate copulas, \cite{Haff2013} shows the consistency and asymptotic normality of the tree-by-tree sequential estimation procedure. With standard arguments~(Taylor expansion and Kullback-Leibler inequality), results in \cite{Haff2013} can be used to show model selection consistency~(under i.i.d. case) of the proposed tree-by-tree sequential selection procedure with BIC when the bivariate copulas of the D-vine are unknown and need to be selected from a fixed number of candidate copulas. Under the conditions of Theorem \ref{MCergodic}, uDvine is stationary and ergodic. Thus, we expect the results in \cite{Haff2013} hold for uDvine with a finite Markov order $p$ and the proposed tree-by-tree sequential selection procedure is consistent. A rigorous theoretical investigation is beyond the scope of this paper and we leave it for future research.}

\subsection{Two-stage MLE for CuDvine}
Given the parametric form of CuDvine, there are three components to be estimated: (a) the marginal distributions $F_1^0(\cdot),\ldots, F_d^0(\cdot)$ of the $d$ component uDvines, (b) the parameters $\bfbeta_1^0,\ldots, \bfbeta_d^0$ of bivariate copulas in the $d$ component uDvines, (c) the parameter $\gamma^0$ of the cross-sectional copula $C(\cdot)$. Throughout this section, we assume that the parametric form~(i.e. the bivariate copula types for each uDvine($p_i$) and the cross-sectional copula type for $C(\cdot)$) of CuDvine is known, and we present the properties of the two-stage MLE under the correct model specification.

Denote $\{\bfy_t=\{y_{ti}\}_{i=1}^d\}_{t=1}^T$ as the observations of the multivariate time series. By differentiating (\ref{CuDvineCDF}), the conditional likelihood function of $\bfy_t$ can be obtained as
\begin{equation}
\begin{split}
f(\bfy_t|\mathcal{F}_{t-1})=c(F_1(y_{t1}|\mathcal{F}_{t-1}^1), \ldots, F_d(y_{td}|\mathcal{F}_{t-1}^d); \gamma)\prod_{i=1}^{d}f_i(y_{ti}|\mathcal{F}_{t-1}^i),
\end{split}
\label{joint}
\end{equation}
where the conditional marginal distributions $F_i(y_{ti}|\mathcal{F}_{t-1}^i)$ and $f_i(y_{ti}|\mathcal{F}_{t-1}^i)$ are defined in Section 2.2.1 and can be derived from the $i$th uDvine($p_i$).

Based on (\ref{joint}), the conditional log-likelihood function is
\begin{align}
\label{loglik}
&L(F_1,\ldots, F_d; \bfbeta_1,\ldots, \bfbeta_d; \gamma \big \vert\{\bfy_t\}_{t=1}^T) =
\sum_{t=p+1}^{T}\log f(\bfy_t|\mathcal{F}_{t-1})\nonumber\\
&=\sum_{t=p+1}^{T}\log c(F_1(y_{t1}|\mathcal{F}_{t-1}^1), \ldots, F_d(y_{td}|\mathcal{F}_{t-1}^d);\gamma)
+ \sum_{i=1}^{d}\sum_{t=p+1}^{T} \log f_i(y_{ti}|\mathcal{F}_{t-1}^i).
\end{align}
The number of parameters to be estimated in (\ref{loglik}) is at least $d\times p$ even if we assume all the bivariate copulas of the uDvines are single-parameter copulas. The full likelihood estimation can be computationally expensive especially when the dimension $d$ is large. To improve computational efficiency, we adapt the two-stage maximum likelihood estimator~(MLE) in the copula literature\citep[e.g.][]{Joe1996a, ChenFan2006a}. The basic idea is to decompose $(\ref{loglik})$ into several components and optimize each component separately.

In the first stage, for $i=1,\ldots, d$, the marginal distribution $F_i^0(\cdot)$ and the parameter $\bfbeta^0_i$ in uDvine$(p_i)$ are estimated using the $i$th component univariate time series $\{y_{ti}\}_{t=1}^T$. Specifically, the marginal distribution $F_i^0(\cdot)$ is estimated by the rescaled empirical distribution function $\hat{F}_i(\cdot)$, where $\hat{F}_i(\cdot) = \frac{1}{T+1}\sum_{t=1}^{T}I(y_{ti}\leq \cdot).$ Given $\hat{F}_i(\cdot)$, the MLE $\hat{\bfbeta}_i$ for $\bfbeta_i^0$ can be calculated by maximizing
\begin{equation}
L_{1i}(\bfbeta_i)=\sum_{t=p+1}^{T}\log f_i(y_{ti}|\mathcal{F}_{t-1}^i)=\sum_{t=p+1}^{T} \left[\log f_i(y_{ti}) + \log w_i(\hat{F}_i(y_{ti}), \cdots, \hat{F}_i(y_{t-p,i}); \bfbeta_i)\right],
\label{1stStageMLE}
\end{equation}
where the last equality follows from (\ref{Dvinepdf2}). \textcolor{black}{Note that maximizing \eqref{1stStageMLE} is equivalent to maximizing $\sum_{t=p+1}^{T} \log w_i(\hat{F}_i(y_{ti}), \cdots, \hat{F}_i(y_{t-p,i}); \bfbeta_i)$ as $\bfbeta_i$ does not affect $f_i(\cdot)$.}

In the second stage, given estimators $\{\hat{F}_i(\cdot)\}_{i=1}^d$ and $\{\hat{\bfbeta}_i\}_{i=1}^d$, the MLE $\hat{\gamma}$ for $\gamma^0$ can be calculated by maximizing
\begin{align}\label{2ndStageMLE}
L_2(\gamma)&=\sum_{t=p+1}^{T}\log c(\hat{F}_1(y_{t1}|\mathcal{F}_{t-1}^1), \ldots, \hat{F}_d(y_{td}|\mathcal{F}_{t-1}^d); \gamma)\\
&=\sum_{t=p+1}^{T}\log c(g_1(\hat{F}_1(y_{t1}), \ldots, \hat{F}_1(y_{t-p,1}); \hat{\bfbeta}_1), \ldots, g_d(\hat{F}_d(y_{td}), \ldots, \hat{F}_d(y_{t-p,d}); \hat{\bfbeta}_d); \gamma),\nonumber
\end{align}
where the last equality follows from (\ref{gfunction}).

\subsection{Consistency and normality of the MLE}
Both the first stage MLE $\{\hat{\bfbeta}_i\}_{i=1}^d$ of parameters $\{\bfbeta_i^0\}_{i=1}^d$ in the uDvines and the second stage MLE $\hat{\gamma}$ of the parameter $\gamma^0$ in the cross-sectional copula are essentially the so-called semiparametric two-stage estimator. A general treatment on its asymptotic properties can be found in \cite{Newey1994}. In the following, we provide the results on consistency and normality for both $\{\hat{\bfbeta}_i\}_{i=1}^d$ and $\hat{\gamma}$ under the context of D-vine based time series.

\subsubsection{Asymptotic properties of $\hat{\bfbeta}_i$}
Given the estimated marginal distribution $\hat{F}_i(\cdot)$, each $\hat{\bfbeta}_i$ is calculated by maximizing the log-likelihood function (\ref{1stStageMLE}). In \cite{ChenFan2006b}, the authors provide asymptotic properties of such two-stage MLE when the univariate time series is generated by a first-order Markov chain based on a bivariate copula. Here, we extend the result to uDvine($p$), which is an arbitrary-order Markov chain based on a D-vine.

Since uDvine($p$) is a generalization of the bivariate copula based first-order Markov chain in \cite{ChenFan2006b}, it is natural to expect that the theoretical properties of $\hat{\bfbeta}_i$ are similar to the ones in \cite{ChenFan2006b}.

\begin{theorem}
	Assume conditions C1-C5 in the supplementary material hold for the $i$th uDvine$(p_i)$, we have $\|\hat{\bfbeta}_i -\bfbeta_i^0\|=o_p(1), $ i.e.  $\hat{\bfbeta}_i$ is consistent.
	\label{1stMLEConsistency}
\end{theorem}

Before stating the result for asymptotic normality, we first introduce some notations for the ease of presentation. Denote $l_i(u_1, \ldots, u_{p+1}; \bfbeta_i)=\log w_i(u_1, \ldots, u_{p+1}; \bfbeta_i)$, $l_{i, \bfbeta}(u_1,\ldots,u_{p+1}; \bfbeta_i)={\partial l_i(u_1, \ldots, u_{p+1}; \bfbeta_i)}/{\partial \bfbeta_i}$, $l_{i, \bfbeta, \bfbeta}(u_1,\ldots,u_{p+1}; \bfbeta_i)={\partial^2 l_i(u_1, \ldots, u_{p+1}; \bfbeta_i)}/{\partial \bfbeta_i \partial \bfbeta_i'}$ and\\
\noindent $l_{i, \bfbeta, k}(u_1,\ldots,u_{p+1}; \bfbeta_i)={\partial^2 l_i(u_1, \ldots, u_{p+1}; \bfbeta_i)}/{\partial \bfbeta_i \partial u_k}$, for $k=1,2,\ldots, p+1$.

Further denote $U_{ti}=F_i^0(Y_{ti})$, $B_i=-E^0(l_{i,\bfbeta,\bfbeta}(U_{ti}, U_{t-1,i},\ldots, U_{t-p,i}; \bfbeta_i^0))$ and
$$A_T^i=\frac{1}{T-p}\sum_{t=p+1}^{T} \left[l_{i, \bfbeta}(U_{ti}, U_{t-1,i},\ldots, U_{t-p,i}; \bfbeta_i^0) + \sum_{k=0}^{p}W_k^i(U_{t-k,i})\right],$$
where $W_k^i(x)=E^0 \left( l_{i,\bfbeta,k+1}(U_{ti}, \ldots, U_{t-p,i};\bfbeta^0_i)(I(x\leq U_{t-k,i})-U_{t-k,i}) \right)$. Define $\Sigma_i = \lim\limits_{T\to\infty}Var^0(\sqrt{T}A_T^i)$.

\begin{theorem}
	Assume conditions A1-A6 in the supplementary material hold for the $i$th uDvine$(p_i)$, we have: (1) $\hat{\bfbeta}_i -\bfbeta_i^0=B_i^{-1}A_T^i + o_p(1/\sqrt{T}) $; (2) $\sqrt{T}(\hat{\bfbeta}_i -\bfbeta_i^0) \to N(0, B_i^{-1}\Sigma_i B_i^{-1})$ in distribution.
	\label{1stMLENormality}
\end{theorem}

As noted in \cite{ChenFan2006b}, the appearance of the extra $p+1$ terms $\{W_k^i\}_{k=0}^p$ in $A_T^i$ is due to the nonparametric estimation of the marginal distribution $F_i^0(\cdot)$, and if $F_i^0(\cdot)$ is known, the terms $\{W_k^i\}_{k=0}^p$ will disappear.

\subsubsection{Asymptotic properties of $\hat{\gamma}$}
Given $\{\hat{F}_i(\cdot)\}_{i=1}^d$ and $\{\hat{\bfbeta}_i\}_{i=1}^d$, $\hat{\gamma}$ can be calculated by maximizing the log-likelihood function \eqref{2ndStageMLE}. Compared to $\hat{\bfbeta}_i$, $\hat{\gamma}$ is obtained based on a log-likelihood function that depends on both the estimated infinite-dimensional functions $\{\hat{F}_i(\cdot)\}_{i=1}^d$ and the extra finite-dimensional estimators $\{\hat{\bfbeta}_i\}_{i=1}^d$. The presence of the extra $\{\hat{\bfbeta}_i\}_{i=1}^d$ is the main difference between the setting of $\hat{\gamma}$ and the setting of $\hat{\bfbeta}_i$. However, the consistency and normality results still hold, with an extra term in the asymptotic covariance due to the presence of $\{\hat{\bfbeta}_i\}_{i=1}^d$.

\cite{ChenFan2006a} provides asymptotic properties of such second stage MLE under the SCOMDY framework, where the component univariate time series follow conditional univariate models such as ARMA and GARCH. As discussed in Section 2.3, CuDvine is constructed via a hybrid modeling approach with the component univariate time series being semiparametric uDvines. This difference makes {parts of} the asymptotic result of $\hat{\gamma}$ for CuDvine distinct from the one in \cite{ChenFan2006a}.

\begin{theorem}
	Assume conditions D and E in the supplementary material hold for CuDvine, we have $\|\hat{\gamma} -\gamma^0\|=o_p(1),$ i.e. $\hat{\gamma}$ is consistent.
	\label{2ndStageMLEConsistency}
\end{theorem}

Given the true marginal distributions $\{{F}_i^0(\cdot)\}_{i=1}^d$ and true uDvine parameters $\{{\bfbeta}_i^0\}_{i=1}^d$, we denote {$F_i(Y_{ti}|\mathcal{F}_{t-1}^i)=g_i(F_i^0(Y_{ti}), \ldots, F_i^0(Y_{t-p,i}); \bfbeta_i^0) = V_{ti},$}
\noindent where $\{(V_{t1}, \ldots, V_{td})\}_{t=1}^T$ can be thought as the unobserved {\it i.i.d.} copula process generated by the cross-sectional copula $C(v_1,\ldots, v_d; \gamma^0)$. Denote $g_{i,\bfbeta}(u_1,\ldots, u_{p+1};\bfbeta_i)=\partial g_i(u_1,\ldots, u_{p+1};\bfbeta_i)/\partial \bfbeta_i$ and $g_{i,k}(u_1,\ldots, u_{p+1};\bfbeta_i)=\partial g_i(u_1,\ldots, u_{p+1};\bfbeta_i)/\partial u_k$ for $k=1,\ldots, p+1$.

We further denote $h(v_1,\ldots, v_d; \gamma)=\log c(v_1,\ldots, v_d; \gamma)$, $h_\gamma(v_1,\ldots, v_d;\gamma)=\partial h(v_1,\ldots, v_d; \gamma)/\partial \gamma$, $h_{\gamma,\gamma}(v_1,\ldots, v_d;\gamma)=\partial^2 h(v_1,\ldots, v_d; \gamma)/\partial \gamma \partial \gamma'$ and
$h_{\gamma,i}(v_1,\ldots, v_d; \gamma)=\partial^2 h(v_1,\ldots, v_d; \gamma)/\partial \gamma \partial v_i$ for $i=1,\ldots, d$. Denote $U_{ti}=F_i^0(Y_{ti})$ and
$$A^*_T = \frac{1}{T-p}\sum_{t=p+1}^{T}\left[h_\gamma(V_{t1},\ldots, V_{td};\gamma^0) + \sum_{i=1}^{d} Q_{\gamma i}(U_{ti})\right] + \sum_{i=1}^{d} B_{\bfbeta}^i B_i^{-1}A_T^i,$$
where $Q_{\gamma i}(x) = E^0\left[ h_{\gamma,i}(V_{t1},\ldots, V_{td},\gamma^0)\sum_{k=0}^{p}g_{i,k+1}(U_{ti}, \ldots, U_{t-p,i}; \bfbeta_i^0)(I(x \leq U_{t-k,i})-U_{t-k,i})\right]$, $B_{\bfbeta}^i = E^0\left[ h_{\gamma,i}(V_{t1},\ldots, V_{td},\gamma^0)g_{i,\bfbeta}(U_{ti}, \ldots, U_{t-p,i}; \bfbeta_i^0)'\right]$
and $B_i^{-1}A_T^i$ are defined in Theorem \ref{1stMLENormality}. Finally, denote $B^*=-E^0(h_{\gamma,\gamma}(V_{t1},\ldots, V_{td};\gamma_0))$ and $\Sigma^*=\lim\limits_{T\to\infty}Var^0(\sqrt{T}A_T^*)$.
\begin{theorem}	\label{2ndStageMLENormality}
	Assume conditions D and N in the supplementary material hold for CuDvine, we have:
	(1) $\hat{\gamma} -\gamma^0={B^*}^{-1}A_T^* + o_p(1/\sqrt{T}) $; (2) $\sqrt{n}(\hat{\gamma} -\gamma^0) \to N(0, {B^*}^{-1}\Sigma^* {B^*}^{-1})$ in distribution.
\end{theorem}
Note that the asymptotic result for $\hat{\gamma}$ is similar to the one for $\hat{\bfbeta}_i$. The extra $d$ terms $\{Q_{\gamma i}\}_{i=1}^d$ are introduced by the nonparametric estimation of the marginal distributions $\{F_i^0(\cdot)\}_{i=1}^d$, and the extra $d$ terms $\{B_{\bfbeta}^iB_i^{-1}A_T^i\}_{i=1}^d$ are introduced by the estimation of the uDvine parameters $\{\bfbeta_i^0\}_{i=1}^d$. As observed in \cite{Newey1994}, the estimation of $\bfbeta_i^0$ does not influence the asymptotic covariance of $\hat{\gamma}$ if $B_{\bfbeta}^i=0$. {In \cite{ChenFan2006a}, there are no $\{B_{\bfbeta}^iB_i^{-1}A_T^i\}_{i=1}^d$ terms in $A_T^*$, due to the conditional modeling approach of the component univariate time series.}

There is no closed form solution for the asymptotic covariance for the second-stage MLE. Though the standard plug-in estimator can be constructed, it will be quite complicated to implement. A practical solution to the estimation of the asymptotic covariance is parametric bootstrap, e.g. see \cite{Zhao2017}.

\section{Numerical Experiments}
\subsection{Flexibility of uDvine}
\subsubsection{\textcolor{black}{Approximating GARCH and GJR-GARCH processes}}
In this section, we demonstrate the flexibility of uDvine in terms of how well it approximates a GARCH~\citep{Bollerslev1986} or GJR-GARCH process~\citep{GLOSTEN1993}. The GARCH process is one of the most widely used univariate time series models in financial markets and is able to capture the unique features observed in stock returns, such as heavy tailedness and volatility clustering. The GJR-GARCH process further introduces asymmetry to the GARCH process by allowing the conditional variance to respond differently to positive and negative stock returns, and it contains the GARCH process as a special case. Specifically, a univariate time series $\{Y_t\}$ follows a GJR-GARCH process, if
\begin{align*}
&Y_{t}=\sigma_{t}\eta_{t},~\eta_{t}\overset{i.i.d.}{\sim} N(0,1),\\
&\sigma^2_{t}=\omega_0+\omega_1 \sigma_{t-1}^2+ \omega_2 Y_{t-1}^2 + \omega_3 I(Y_{t-1}>0).
\end{align*}
If $\omega_3 \equiv 0$, then $\{Y_t\}$ reduces to a GARCH process. We set the parameters to be $[\omega_0, \omega_1, \omega_2, \omega_3]=[0.05, 0.85, 0.1, 0]$ for the GARCH process and $[\omega_0, \omega_1, \omega_2, \omega_3]=[0.05, 0.85, 0.1, 0.05]$ for the GJR-GARCH process. According to \cite{OhPatton2013}, the parameters broadly match the values of estimation from the real world financial data.

We use uDvine to model $\{Y_t\}_{t=1}^T$ simulated from the above GARCH or GJR-GARCH process. We do not fix the parametric form of the uDvine but instead use the sequential selection method in Section 3.1 to build the uDvine in a data-driven fashion. This is different from \cite{LoaizaMaya2017} where the authors fix the parametric forms of vine-copula beforehand. The candidate pool for the bivariate copulas consists of 40 different bivariate copulas that are implemented in the R package \texttt{VineCopula}~\citep[][]{Schepsmeier2017}. We assess the goodness of approximation by the out-of-sample performance on predicting one-day ahead conditional Value at Risk~(VaR) for $Y_t$. Conditional VaR is the most commonly used extreme risk measure in financial applications. For $0<q<1$, VaR$^{1-q}_{t}$ is defined as the $1-q$ conditional quantile of $Y_t$ given the past information $\mathcal{F}_{t-1}=\sigma(Y_{t-1},Y_{t-2},\ldots)$, where $q$ is usually taken to be 0.1 or 0.05. Note that extreme quantile tracking is never an easy task, especially when the underlying time series has complicated behavior such as heavy-tailedness, volatility clustering and asymmetric nonlinear dependence.

Specifically, we first fit the uDvine based on a training set $\{Y_t\}_{t=1}^{T_1}$. Then using the fitted uDvine, we calculate the one-day ahead conditional VaR$_t^{1-q}$ for each $Y_t$ in the test set $\{Y_t\}_{t=T_1+1}^{T_1+T_2}$. The one-day ahead VaR$_t^{1-q}$ is calculated based on 1000 bootstrapped samples from the fitted uDvine. The detailed algorithm for generating bootstrapped samples from uDvine can be found in Section \S5.1 of the supplementary material. The true $\{Y_t\}_{t=T_1+1}^{T_1+T_2}$ are then compared with the $\{\text{VaR}_t^{1-q}\}_{t=T_1+1}^{T_1+T_2}$ and the number of violations are recorded. A violation happens when the observed $Y_t$ is larger than the corresponding VaR$_t^{1-q}$ given by the uDvine. If uDvine approximates the GARCH or GJR-GARCH process well, the expected proportion of violations in the test set should be close to $q$.

We set $T_1=1000, 2000, 5000$, $T_2=100$ and $q^{0}=0.1, 0.05$. For each combination of $(T_1, T_2, q^0)$, we repeat the experiment 500 times. The $i$th experiment gives a realized violation percentage $q_i$ and we report the average percentage,
$\bar{q}=\sum_{i=1}^{500}q_i/500$, in Table \ref{VARonGJR} for both the GARCH and GJR-GARCH process. We also report in the table the $p$-values for testing $E(q_i)=q^0$ using one-sample $Z$-tests based on the observed $\{q_i\}_{i=1}^{500}$.

\begin{table}[h!]
	\centerline{
		\begin{tabular}{ccccc|cccc}\hline\hline
			\multicolumn{5}{c}{GARCH} & \multicolumn{4}{c}{GJR-GARCH}\\\hline
			$T_1$ & $\bar{q}(q^0=0.1)$ & $p$-value & $\bar{q}~(q^0=0.05)$ & $p$-value & $\bar{q}~(q^0=0.1)$ & $p$-value & $\bar{q}~(q^0=0.05)$ & $p$-value  \\\hline
			1000 &  0.106 &  0.001 & 0.055 & 0.000 &  0.107 &  0.001 & 0.056 & 0.000 \\\hline
			2000 &  0.103 &  0.176 & 0.052 & 0.196 & 0.104 &  0.125 & 0.052 & 0.319 \\\hline
			5000 &  0.102 &  0.267 & 0.051 & 0.468 & 0.104 &  0.133 & 0.053 & 0.170 \\\hline\hline
	\end{tabular}}
	\caption{ {\it The performance of uDvine on approximating the one-day ahead conditional VaR for the GARCH and GJR-GARCH processes.}}
	\label{VARonGJR}
\end{table}

As observed from Table \ref{VARonGJR}, for all combinations of $(T_1, T_2, q^0)$, the average violation percentage $\bar{q}$ achieved by uDvine is very close to the target level $q^0$, for both the GARCH and GJR-GARCH process. In addition, it passes the $Z$-test when the training set is large enough~($T_1\geq 2000$). {For $T_1=5000$, under both GARCH and GJR-GARCH process, we find that around 95\% of the uDvines are selected to be uDvine(1) with a $t$-copula and around 5\% are selected to be uDvine(2) with two $t$-copulas. This matches the analytic findings of Example 3 in Section \S1 of the supplementary material.}

\subsubsection{\textcolor{black}{Approximating higher-order AR processes}}
\textcolor{black}{In this section, we demonstrate the flexibility of uDvine in terms of how well it approximates a higher-order autoregressive~(AR) process. Specifically, a stationary AR(9) process $\{Y_t\}$ is generated via
\begin{align*}
	Y_{t} = & 0.7Y_{t-1}-0.6Y_{t-2}+0.6Y_{t-3}-0.5Y_{t-4}+0.5Y_{t-5}-0.5Y_{t-6}\\ + & 0.6Y_{t-7}-0.4Y_{t-8}+0.4Y_{t-9} +\epsilon_t, ~ \epsilon_t\overset{i.i.d.}{\sim} N(0,1).
\end{align*}
Note that $\{Y_t\}$ is a Markov chain of order 9. Though a straightforward model, an AR(9) process is not easy to approximate due to its high autoregressive order.}

\textcolor{black}{We use uDvine to model $\{Y_t\}_{t=1}^T$ simulated from the above AR(9) process. Same as in Section 4.1.1, we do not fix the parametric form of the uDvine but instead use the sequential selection method in Section 3.1 to build the uDvine in a data-driven fashion. The candidate pool for the bivariate copulas consists of 40 different bivariate copulas that are implemented in the R package \texttt{VineCopula}~\citep[][]{Schepsmeier2017}. We assess the goodness of approximation by the performance on out-of-sample one-day ahead prediction for $Y_t$ given the past information $\mathcal{F}_{t-1}=\sigma(Y_{t-1},Y_{t-2},\ldots)$.}

\textcolor{black}{Specifically, we first fit the uDvine based on a training set $\{Y_t\}_{t=1}^{T_1}$. Then using the fitted uDvine, we calculate the one-day ahead prediction for each $Y_t$ given $\mathcal{F}_{t-1}$ in the test set $\{Y_t\}_{t=T_1+1}^{T_1+T_2}$. The one-day ahead prediction $\hat{\mu}_t$ is calculated as the sample mean of 1000 bootstrapped samples from the fitted uDvine. The detailed algorithm for generating bootstrapped samples from uDvine can be found in Section \S5.1 of the supplementary material. For comparison, we consider the oracle one-day ahead prediction with $\mu_t=E(Y_t|\mathcal{F}_{t-1})=0.7Y_{t-1}-0.6Y_{t-2}+0.6Y_{t-3}-0.5Y_{t-4}+0.5Y_{t-5}-0.5Y_{t-6}+  0.6Y_{t-7}-0.4Y_{t-8}+0.4Y_{t-9}$ based on the AR(9) process. The true $\{Y_t\}_{t=T_1+1}^{T_1+T_2}$ are compared with $\{\hat{\mu}_t\}_{t=T_1+1}^{T_1+T_2}$ or $\{{\mu}_t\}_{t=T_1+1}^{T_1+T_2}$ via mean squared error~(MSE) $\frac{1}{T_2}\sum_{t=T_1+1}^{T_1+T_2}(Y_t-\hat{\mu}_t)^2$ or $\frac{1}{T_2}\sum_{t=T_1+1}^{T_1+T_2}(Y_t-{\mu}_t)^2$.}

\textcolor{black}{We set $T_1=200, 500, 1000$ and $T_2=50$. For each combination of $(T_1, T_2)$, we repeat the experiment 500 times. We report the mean and median MSE across the 500 experiments in Table \ref{AR9} for both the oracle and uDvine prediction. In addition, we report the mean selected order $\bar{p}$ of uDvine across the 500 experiments. As can be seen, the performance of uDvine improves as the sample size $T_1$ increases and is comparable to the oracle prediction. The selected uDvine order is close to 9, which is the true order of the Markov chain.}

\begin{table}[h!]
	\textcolor{black}{
	\centerline{
		\begin{tabular}{cccc|cc}\hline\hline
			\multicolumn{4}{c}{uDvine} & \multicolumn{2}{c}{Oracle}\\\hline
			$T_1$ & mean MSE & med MSE& $\bar{p}$ & mean MSE & med MSE  \\\hline
			200 & 1.288 & 1.232 & 8.16 & 1.023 & 1.003 \\
			500 & 1.101 & 1.060 & 8.86 & 0.996 & 1.001 \\
			1000 & 1.036 & 1.019 & 9.11 & 0.990 & 0.978 \\\hline\hline
	\end{tabular}}
	\caption{ {\it \color{black} The performance of uDvine on approximating an AR(9) process in terms of one-day ahead prediction error.}}
	\label{AR9}}
\end{table}

\subsection{Performance of the sequential selection for uDvine}
In this section, we investigate the performance of the tree-by-tree sequential selection procedure described in Section 3.1. Specifically, we conduct numerical experiments for three uDvine(2)s with different parameter settings. The marginal distributions for all uDvine(2)s are set to be $N(0,1)$.

For the first uDvine(2), we set tree 1 to be Gaussian($\rho^1=0.7$) copula and tree 2 to be Gumbel($\alpha^1=1.25$) copula. For the second uDvine(2), we set tree 1 to be $t_{\nu^2=3}(\rho^2=0.7)$ copula and tree 2 to be Clayton$(\theta^2=0.5)$ copula. For the third uDvine(2), we set tree 1 to be Gaussian($\rho^{3}_1=0.7$) copula and tree 2 to be Gaussian($\rho^{3}_2=0.3$) copula. The parameters of all the bivariate copulas are specified to make the Kendall's tau of tree 1 to be 0.5 and that of tree 2 to be 0.2.

We assume the candidate pool of bivariate copulas to be (Gaussian, $t$, Clayton, Gumbel, Frank, Joe), which contains the most widely used copulas in practice. For each uDvine(2), we perform the sequential selection procedure under sample size of $T=1000, 2000$ and $5000$. For each sample size $T$, we repeat the numerical experiment 500 times. We report the percentage of correctly selected order of the uDvine and the percentage of correctly selected copulas for each tree of the uDvine. The results are displayed in Table \ref{treebytree}. As suggested by the table, the sequential selection procedure performs well in both order selection and copula selection. Also, the performance is improving with the increase of sample size $T$.

\begin{table}[h]
	\centering
	\begin{tabular}{cccc}\hline\hline
		$T$    & $\text{order $p=2$}$   & $\text{tree 1 (Gaussian)}$  & $\text{tree 2 (Gumbel)}$    \\\hline
		1000 & 0.99 & 0.99 & 0.88\\
		2000 & 0.98 & 0.97 & 0.97 \\
		5000 & 1.00 & 0.99 & 1.00 \\\hline
		$T$    & $\text{order $p=2$}$   & $\text{tree 1 ($t_3$)}$  & $\text{tree 2 (Clayton)}$    \\\hline
		1000 & 0.98 & 0.98 & 0.97\\
		2000 & 1.00 & 1.00 & 1.00\\
		5000 & 0.99 & 1.00 & 1.00 \\\hline
		$T$    & $\text{order $p=2$}$   & $\text{tree 1 (Gaussian)}$  & $\text{tree 2 (Gaussian)}$    \\\hline
		1000 & 0.99 & 0.99 & 0.92\\
		2000 & 1.00 & 1.00 & 0.98\\
		5000 & 1.00 & 0.99 & 1.00 \\\hline\hline
	\end{tabular}
	\caption{\it Performance of the tree-by-tree sequential selection procedure for three different uDvine(2).}
	\label{treebytree}
\end{table}

\subsection{Performance of the two-stage MLE for CuDvine}
In this section, we investigate the finite-sample performance of the two-stage MLE for a three-dimensional CuDvine consisting of the three uDvine(2) in Section 4.2. To fully specify CuDvine, we set the cross-sectional copula $C(\cdot)$ to be Gaussian with $(\rho_{12}, \rho_{13}, \rho_{23})=(0.2, 0.5, 0.8)$. We assume that the parametric form~(i.e. the bivariate copula types for each uDvine($2$) and the cross-sectional copula type) of CuDvine is known.

We study the performance of the two-stage MLE under sample size $T=1000, 2000$ and $5000$. For each sample size $T$, we repeat the experiment 500 times. Table \ref{MLE1} summarizes the results, which show the two-stage MLE is consistent and the accuracy of MLE is improving with $T$ growing.

\begin{table}[h]
	\centering
	\begin{tabular}{cccccc}\hline\hline
		$T$    & $\rho^1=0.7$   & $\alpha^1=1.25$  &  $\rho^2=0.7$ & $\nu^2=3$ &  $\theta^2=0.5$  \\\hline
		1000 & 0.699(0.030) & 1.250(0.035) & 0.694(0.034) & 3.374(0.760) & 0.482(0.088)\\
		2000 & 0.700(0.024) & 1.248(0.024) & 0.700(0.022) & 3.146(0.558) & 0.489(0.068)\\
		5000 & 0.700(0.016) & 1.247(0.015) & 0.699(0.016) & 3.090(0.299) & 0.495(0.041)\\\hline
		$T$    & $\rho^{3}_1=0.7$   & $\rho^{3}_2=0.3$  & $\rho_{12}=0.2$ &  $\rho_{13}=0.5$ & $\rho_{23}=0.8$  \\\hline
		1000 & 0.692(0.026) & 0.300(0.032) & 0.202(0.032) & 0.498(0.027) & 0.795(0.012) \\
		2000 & 0.699(0.021) & 0.296(0.019) & 0.198(0.024) & 0.498(0.018) & 0.796(0.010) \\
		5000 & 0.700(0.013) & 0.301(0.012) & 0.201(0.013) & 0.499(0.011) & 0.799(0.005) \\\hline\hline
	\end{tabular}
	\caption{\it Performance of the two-stage MLE for a three-dimensional CuDvine. The sample standard deviations of the MLE are in brackets.}
	\label{MLE1}
\end{table}

\subsection{\textcolor{black}{Performance of CuDvine for high-dimensional time series}}
\textcolor{black}{In this section, we demonstrate the ability of CuDvine to model high-dimensional time series and to track large dynamic covariance matrices. Specifically, we generate a 100-dimensional time series via the multivariate GARCH-CCC~(constant conditional correlation) process in \cite{Bollerslev1990} and model its behavior via CuDvine. A $d$-dimensional multivariate time series $\{Y_{ti}\}_{i=1}^d$ follows a GARCH-CCC process if
\begin{align*}
&Y_{ti}=\sigma_{ti}\eta_{ti}, ~ \sigma^2_{ti}=\omega_{i0}+\omega_{i1} \sigma_{t-1,i}^2+ \omega_{i2} Y_{t-1,i}^2, \text{ for } i =1,2,\cdots, d,\\
& \bfeta_{t}=(\eta_{t1}, \eta_{t2},\cdots, \eta_{td}) \overset{i.i.d.}{\sim} E(\eta_{ti})=0, Var(\eta_{ti})=1, Cov(\bfeta_t)=R.
\end{align*}
Marginally each univariate time series $\{Y_{ti}\}, i=1,2,\cdots, d$ follows a GARCH(1,1) process with conditional variance $\sigma_{ti}^2.$ Denote $D_t=diag(\sigma_{t1}^2, \sigma_{t2}^2, \cdots, \sigma_{td}^2)$, the conditional covariance matrix of $\bfY_t=(Y_{t1},\cdots,Y_{td})$ given past information $\mathcal{F}_{t-1}$ is $\Sigma_t=D_t^{1/2}R D_t^{1/2}$.}

\textcolor{black}{For each univariate GARCH process, we set $(\omega_{i0}, \omega_{i1}, \omega_{i2})=(0.05, 0.85, 0.1)$ for $i=1,\cdots, d$ as in Section 4.1.1. To fully specify the GARCH-CCC process, we need to set the distribution of $\bfeta_t$. To resemble financial data, we set $\bfeta_t$ to follow a multivariate $t$-distribution with degree of freedom $\nu=6$. As for the correlation matrix $R$ of $\bfeta_t$, we use the block factor structure discussed in Section 2.3.1. Specifically, we set $d=100$, $m=4$ and $(d_1,d_2,d_3,d_4)=(25,25,25,25)$, i.e. the multivariate time series is of dimension 100 with four blocks each having 25 time series. We set $\bfphi_0=(\phi_{10},\phi_{20},\phi_{30},\phi_{40})=(1,1,1.2,1.2)$, $\bfphi_1=(\phi_{11},\phi_{21},\phi_{31},\phi_{41})=(0.8,0.8,1,1)$, implying within-block correlation of $0.62, 0.71$ and between-block correlation of $0.38, 0.40, 0.42$~(see equation \eqref{blockfactor}). Note that the multivariate GARCH-CCC process implies that given $\mathcal{F}_{t-1}$, the conditional cross-sectional dependence of $\bfY_t$ follows a $t$-copula with degree of freedom $\nu$ and correlation matrix $R$.}

\textcolor{black}{We use CuDvine to model/approximate the high-dimensional time series $\{\bfY_t\}_{t=1}^T$ simulated from the above multivariate GARCH-CCC process. For each univariate time series $\{Y_{ti}\}_{t=1}^T, i=1,2,\cdots, 100$, the uDvine is estimated in the same fashion as in Section 4.1.1. The cross-sectional copula of CuDvine is set to be $t$-copula with the block factor structure. The parameter $(\bfphi_0, \bfphi_1, \nu)$ of the $t$-copula is estimated via two-stage MLE. We assess the performance of CuDvine on modeling high-dimensional time series by its out-of-sample prediction of the conditional covariance matrix $\Sigma_t=D_t^{1/2}R D_t^{1/2}$ of $\bfY_t$ given $\mathcal{F}_{t-1}$. Note that an accurate prediction of $\Sigma_t$ requires a precise estimation of both the marginal variance $D_t$ and the high-dimensional correlation matrix $R$.}

\textcolor{black}{Specifically, we first fit the CuDvine based on a training set $\{\bfY_t\}_{t=1}^{T_1}$. Then using the fitted CuDvine, we calculate the conditional covariance matrix for each $\bfY_t$ given $\mathcal{F}_{t-1}$ in the test set $\{\bfY_t\}_{t=T_1+1}^{T_1+T_2}$. The estimated conditional covariance $\widehat{\Sigma}_t$ of day $t$ is computed as the sample covariance of 1000 bootstrapped samples $\{\bfY_t^b\}_{b=1}^{1000}$ from the fitted CuDvine given $\mathcal{F}_{t-1}$. The detailed algorithm for generating bootstrapped samples from CuDvine can be found in Section \S5.2~(Scenario A) of the supplementary material. We compare $\widehat{\Sigma}_t$ with the true covariance matrix $\Sigma_t=D_t^{1/2}R D_t^{1/2}$ by calculating the mean scaled Frobenius norm of error~(MFE) $\frac{1}{T_2}\sum_{t=T_1+1}^{T_1+T_2}\|\widehat{\Sigma}_t-{\Sigma}_t\|^2/\|\Sigma_t\|^2$.}

\textcolor{black}{We set $T_1=1000, 2000, 5000$ and $T_2=50$. For each combination of $(T_1, T_2)$, we repeat the experiment 500 times. We report the mean and median MFE across the 500 experiments in Table \ref{GARCHCCC}. We also report the performance of two-stage MLE for $(\phi_{10}, \phi_{11}, \nu)$ across the 500 experiments~(Estimation for the rest $\bfphi_0, \bfphi_1$ is similar and is omitted to save space).}

\textcolor{black}{As can be seen, CuDvine can track the dynamics of the large conditional covariance matrix $\Sigma_t$ accurately, making around only 4\% to 9\% relative error, confirming the ability of CuDvine to model high-dimensional time series. The estimated $(\hat{\bfphi}_0, \hat{\bfphi}_1, \hat{\nu})$ of the $t$-copula is close to the true parameter value, despite the fact that CuDvine is a misspecified model for the multivariate GARCH-CCC process.}

\begin{table}[h!]
	\textcolor{black}{
		\centerline{
			\begin{tabular}{cccccc}\hline\hline
				$T_1$ & mean MFE & med MFE& $\hat{\phi}_{10}$ & $\hat{\phi}_{11}$ & $\hat{\nu}$  \\\hline
				1000 & 0.091 & 0.085 & 1.018~(0.035) & 0.786~(0.027) & 7.348~(0.551) \\				
				2000 & 0.082 & 0.080 & 1.023~(0.028) & 0.791~(0.020) & 7.342~(0.344) \\
				5000 & 0.046 & 0.045 & 1.021~(0.017) & 0.795~(0.013) & 7.091~(0.231)\\\hline\hline
		\end{tabular}}
		\caption{{\it \color{black} The performance of CuDvine on tracking conditional covariance matrix. The sample standard deviations of the MLE are in brackets.}}
		\label{GARCHCCC}}
\end{table}

\section{Real Data Applications}
In this section, we compare the performance of CuDvine with the vector autoregressive model~(VAR) on the Australian National Electricity Market~(NEM) price dataset\footnote{The data are available freely from https://www.aemo.com.au/Electricity/National-Electricity-Market-NEM/Data-dashboard}. Additional applications of CuDvine in modeling spatio-temporal dependence can be found in Section \S7 of the supplementary material, where improvement of CuDvine over spatial Gaussian model is observed.

\begin{figure}[h]
	\begin{subfigure}{.48\textwidth}
		\centering
		\includegraphics[width=\textwidth]{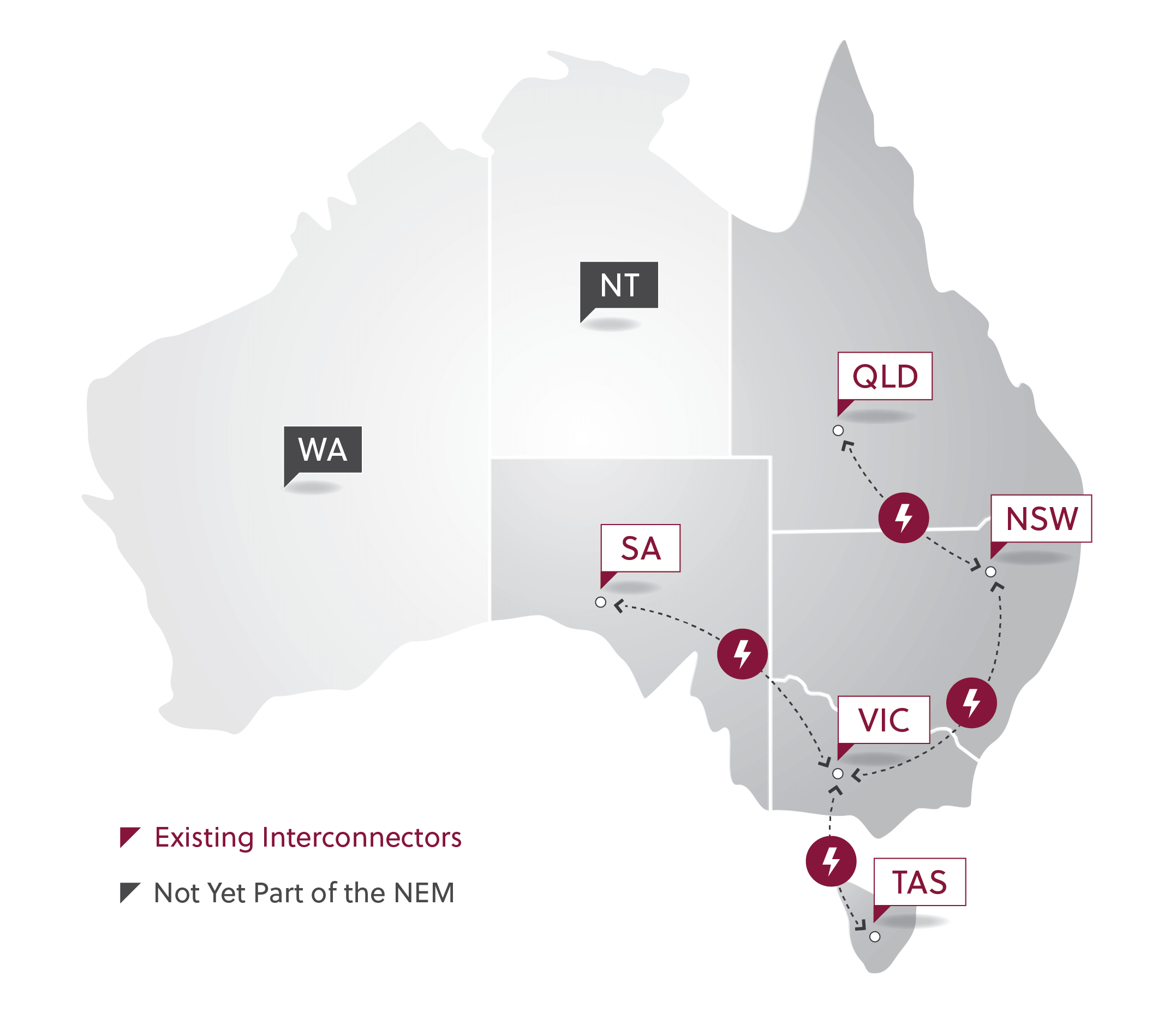}
		\caption{}
		\label{NEM}
	\end{subfigure}%
	\begin{subfigure}{.48\textwidth}
		\centering
		\includegraphics[width=\textwidth, angle=270]{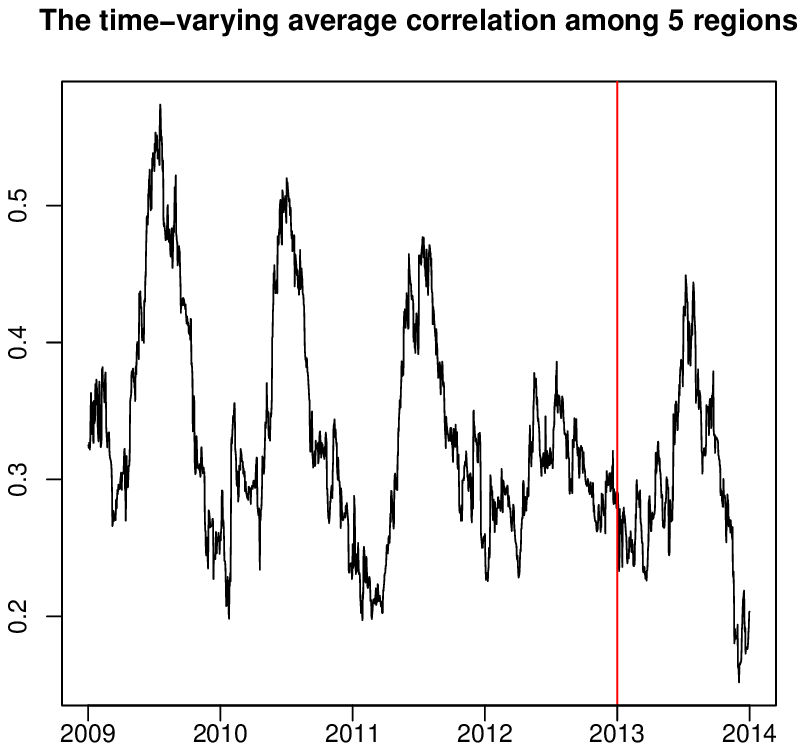}
		\vspace{-0.9cm}
		\caption{}
		\label{TimeVaryingCor}
	\end{subfigure}
	\caption{{\it (a) The locations of the five regions in the Australian National Electricity Market. The dashed lines represent high voltage interconnectors among different regions. (b) The time-varying average correlation across all five regions estimated by the time-varying $t$-copula.}}
	\label{Electricity}
\end{figure}

The NEM interconnects five regional markets of Australia – New South Wales~(NSW), Victoria~(VIC), Queensland~(QLD), Tasmania~(TAS) and South Australia~(SA). Western Australia~(WA) and Northern Territory~(NT) are not connected to the NEM. A map of the relative locations of the regions can be found in Figure \ref{Electricity}(a). Out of the five regions, NSW, VIC and QLD are the major electricity markets with average daily demands of N$_d=8235$, V$_d=5476$ and Q$_d=5913$ megawatts~(MW), while TAS and SA are significantly smaller markets with demands of T$_d=1120$ and S$_d=1441$ MW respectively.

The dataset contains five-year observations of daily maximum electricity price~\textcolor{black}{(in log scale)} of the five regions from 2009-01-01 to 2013-12-31. The day of week effect is removed by a linear regression with seven dummy variables. The ``Seasonal and Trend decomposition using Loess"~(STL) method in \cite{Cleveland1990} is employed to remove the remaining trend and seasonality of each component univariate time series. We train CuDvine and VAR using four-year data from 2009-01-01 to 2012-12-31 (with 1460 days) and hold out the rest one-year data as the test set.

\begin{figure}[h]
	\begin{subfigure}{1\textwidth}
		\centering
		\centerline{\includegraphics[width=0.33\textwidth, angle=270]{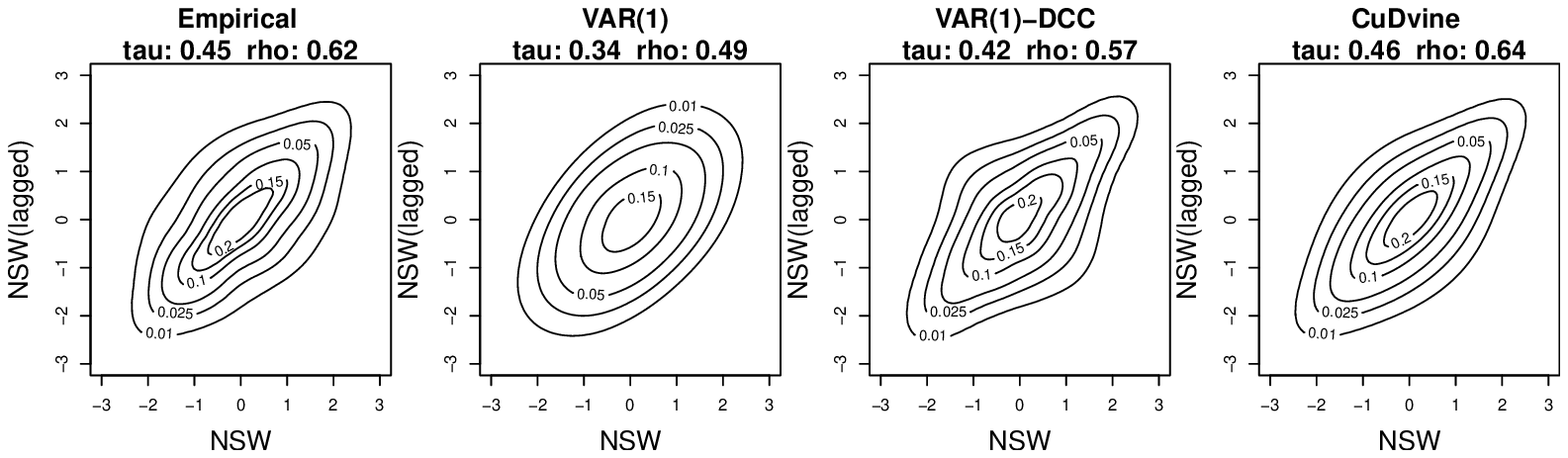}}
		\vspace{-0.4cm}
		\caption{}
	\end{subfigure}%

	\begin{subfigure}{1\textwidth}
		\centering
		\centerline{\includegraphics[width=0.33\textwidth, angle=270]{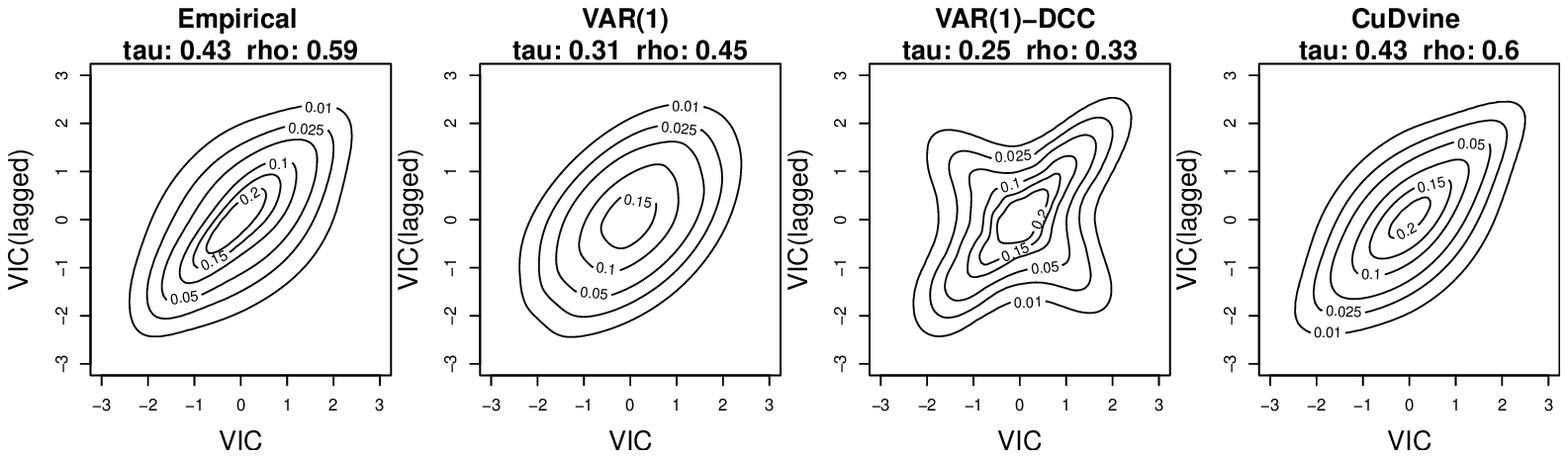}}
		\vspace{-0.4cm}
		\caption{}
	\end{subfigure}%

	\begin{subfigure}{1\textwidth}
		\centering
		\centerline{\includegraphics[width=0.33\textwidth, angle=270]{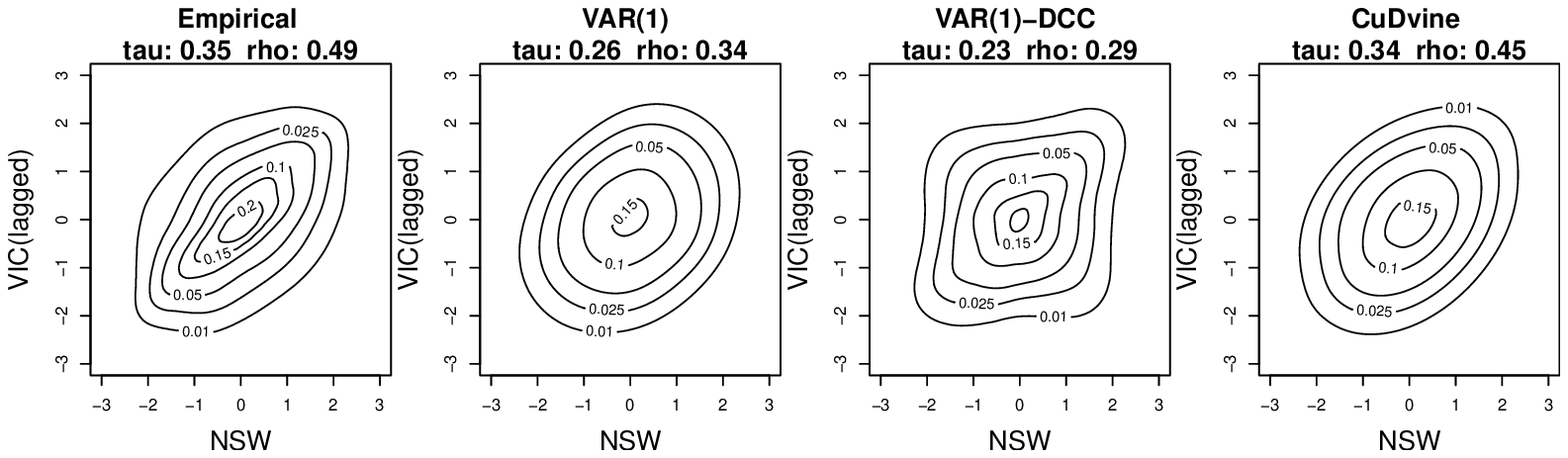}}
		\vspace{-0.4cm}
		\caption{}
    \end{subfigure}
	\caption{{\it \color{black} Contour plot of estimated bivariate copula density (with standard normal margins) for (a) NSW (self lagged) (b) VIC (self lagged) (c) NSW (lagged) v.s. VIC}}
	\label{NSW_VIC_Copula}
\end{figure}

For all five component univariate time series, a uDvine($2$) is selected according to the tree-by-tree sequential selection procedure. For NSW and QLD, a $t$-copula is selected for both tree 1 and tree 2. For VIC and SA, a $t$-copula and a Gumbel copula are selected for tree 1 and tree 2, respectively. For TAS, a BB8 copula is selected for both tree 1 and tree 2. \textcolor{black}{Note that most of the copulas of uDvines are selected as non-Gaussian copulas with tail dependence, indicating potential complicated dependence structure of the data. We further demonstrate this point in the last part of this section, see Figure \ref{NSW_VIC_Copula} later for more details. }

For the cross-sectional dependence, to capture any seasonality in strength of dependence\footnote{{Note that STL only removes the seasonality for each univariate time series, but cannot remove the seasonality in the cross-sectional dependence.}}, we use a 5-dimensional time-varying $t$-copula, where the correlation matrix is designed to evolve according to the DCC model in \cite{Engle2002}~(see Section \S6.1 of the supplementary material for more details). The estimated degree of freedom is 12.79 and the average estimated correlation matrix over the training set is reported in Table \ref{tcopula}.

\begin{table}[]
	\centering
	\begin{tabular}{cccccc}\hline\hline
		& NSW & VIC   & QLD   & TAS   & SA    \\\hline
		NSW & 1   & 0.587~(0.122) & 0.430~(0.118) & 0.278~(0.104) & 0.381~(0.094)\\
		VIC &     & 1     & 0.310~(0.146) & 0.376~(0.098) & 0.566~(0.115) \\
		QLD &     &       & 1     & 0.154~(0.091) & 0.190~(0.111) \\
		TAS &     &       &       & 1     & 0.220~(0.108) \\
		SA  &     &       &       &       & 1    \\\hline\hline
	\end{tabular}
	\caption{{\it The average estimated correlation matrix of the cross-sectional time-varying $t$-copula over the training set period. The standard deviation of each time-varying correlation over the training set period is reported in the brackets.}}
	\label{tcopula}
\end{table}

As shown in Figure \ref{Electricity}(a), there are high voltage interconnectors between NSW and VIC, NSW and QLD, VIC and SA, and VIC and TAS. This pattern matches the estimated parameters of the cross-sectional $t$-copula in Table \ref{tcopula}. The average correlations of the four pairs are respectively 0.587, 0.430, 0.566 and 0.376, which are the highest correlations among all pairs. {For demonstration purpose, we plot the time-varying average correlation across all five regions estimated by the time-varying $t$-copula in Figure \ref{Electricity}(b), which shows strong evidence of seasonality and achieves peak correlation during winter time in Australia.}

\textcolor{black}{The VAR is specified according to AIC where a VAR(1) model is selected. A VAR(2) model is also implemented to investigate the effect of time lags on prediction. For a fair comparison, we also fit a VAR(1)-DCC model, where similar to the time-varying $t$-copula of CuDvine, the covariance matrix of the noise term in VAR(1) evolves based on the DCC model in \cite{Engle2002}~(see Section \S6.2 of the supplementary material for more details).}

We test the model performance on the one-day ahead prediction for each component univariate time series~(NSW, VIC, QLD, TAS, SA), the one-day ahead prediction for the difference between pairs of time series~(VIC-NSW, QLD-NSW, TAS-NSW, SA-NSW, QLD-VIC, TAS-VIC, SA-VIC, TAS-QLD, SA-QLD, SA-TAS), and the one-day ahead prediction for the demand-weighted price of all five time series. On day $t$, denote the price for the five regions as NSW$_t$, VIC$_t$, QLD$_t$, TAS$_t$ and SA$_t$, and the demand-weighted price is defined to be the demand-normalized average price over the five regions
$$\text{(N$_d\cdot$NSW$_t$+V$_d\cdot$VIC$_t$+Q$_d\cdot$QLD$_t$+T$_d\cdot$TAS$_t$+S$_d\cdot$SA$_t$)/(N$_d$+V$_d$+Q$_d$+T$_d$+S$_d$).}$$
Note that the demand-weighted price can be potentially used as a price-index of the Australian National Electricity Market.

For each day in the test set, we generate the one-day ahead prediction distribution based on 1000 bootstrapped samples from \textcolor{black}{the fitted CuDvine, VAR(1), VAR(2) and VAR(1)-DCC model}. The detailed algorithm for generating bootstrapped samples from CuDvine can be found in Section \S5.2~(Scenario A) of the supplementary material. To evaluate the performance of prediction, we consider two out-of-sample metrics, CRPS and QRPS, see \cite{Gneiting2007}. CRPS is a metric for overall prediction accuracy and QRPS is a metric for prediction accuracy of a specific quantile~(e.g. 95\% quantile). Smaller CRPS and QRPS indicate better prediction. {For each day $t$ in the test set, we calculate the CRPS$_t$ and QRPS$_t$ for the fitted CuDvine and VAR models respectively, based on the true observation and the bootstrapped prediction distribution.}

The average CRPS\footnote{The average CRPS/QRPS is defined as the sample average of the CRPS/QRPS's achieved by CuDvine/VAR for each day over the entire test set.} of one-day ahead prediction for NSW, VIC, QLD, TAS and SA achieved by \textcolor{black}{CuDvine, VAR(1), VAR(2) and VAR(1)-DCC} are presented in Table \ref{MarginalCRPS}. We also report the percentage of days in the test set when the CRPS of CuDvine is better than that of the VAR(1)-DCC model, \textcolor{black}{as VAR(1)-DCC gives the best performance among the three VAR variants}. In terms of CRPS, CuDvine outperforms VAR(1)-DCC in every time series around two thirds of the days in the test set and always gives the best overall performance among the four models. We report the average CRPS of one-day ahead prediction for the difference between pairs of time series in Table \ref{DiffCRPS}. It is consistently shown that CuDvine is superior to \textcolor{black}{the three VAR variants} in modeling the difference between pairs.

\begin{table}[]
	\centering
	\begin{tabu}{cccccc}\hline\hline
		& NSW       & VIC       & QLD     &TAS &SA \\\hline
		CuDvine    & 0.150 & 0.171 & 0.368 & 0.230 & 0.351   \\
        VAR(1)      & 0.171 & 0.187 & 0.408 & 0.253 & 0.372 \\
        \rowfont{\color{black}}
		VAR(2)      & 0.172 & 0.187 & 0.407 & 0.252 & 0.371 \\
		\rowfont{\color{black}}
		VAR(1)-DCC  & 0.161 & 0.183 & 0.383 & 0.248 & 0.365  \\
		\rowfont{\color{black}}
		Percentage    & 66.30\% & 67.12\% & 70.41\% & 69.86\% & 68.22\% \\\hline\hline
	\end{tabu}
	\caption{\it Average CRPS for CuDvine and three VAR variants, and the percentage of days that CuDvine is better than VAR(1)-DCC for each component univariate time series.}
	\label{MarginalCRPS}
\end{table}

\begin{table}[]
	\centering
	\begin{tabu}{cccccc}\hline\hline
		CRPS           & VIC-NSW   & QLD-NSW & TAS-NSW  & SA-NSW & QLD-VIC    \\\hline
		CuDvine   & 0.153 & 0.369 & 0.256 & 0.352 & 0.385 \\
		VAR(1)     & 0.205 & 0.425 & 0.289 & 0.371 & 0.432 \\
		\rowfont{\color{black}}
		VAR(2)     & 0.205 & 0.422 & 0.290 & 0.369 & 0.431 \\
		\rowfont{\color{black}}
		VAR(1)-DCC     & 0.193 & 0.399 & 0.282 & 0.364 & 0.411 \\
		\rowfont{\color{black}}
		Percentage     & 81.37\% & 75.07\% & 71.78\% & 67.12\% & 72.05\% \\\hline
		CRPS           & TAS-VIC & SA-VIC & TAS-QLD & SA-QLD & SA-TAS   \\\hline
		CuDvine   & 0.239 & 0.302 & 0.449 & 0.554 & 0.399  \\
		VAR(1)    & 0.280 & 0.338 & 0.492 & 0.585 & 0.426 \\
		\rowfont{\color{black}}
		VAR(2)    & 0.281 & 0.338 & 0.487 & 0.583 & 0.426 \\
		\rowfont{\color{black}}
		VAR(1)-DCC    & 0.275 & 0.338 & 0.471 & 0.570 & 0.421 \\
		\rowfont{\color{black}}
		Percentage     & 74.52\% & 74.25\% & 70.96\% & 67.12\% & 68.49\% \\\hline\hline		
	\end{tabu}
	\caption{\it Average CRPS for CuDvine and three VAR variants, and the percentage of days that CuDvine is better than VAR(1)-DCC for the difference between pairs of time series.}
	\label{DiffCRPS}
\end{table}

We present the prediction result for the demand-weighted price in Table \ref{SumCRPS}. We report the average CRPS and the average QRPS of the 95\% quantile. \textcolor{black}{CuDvine delivers the best performance in both metrics while VAR(1)-DCC comes second.} Based on the bootstrapped prediction distribution, for each day in the test set, we also construct one-day ahead 95\% prediction interval~(P.I.) and 95\% Value at Risk~(VaR) for the demand-weighted price. We present the empirical coverage rates\footnote{The empirical coverage rate of P.I. is defined to be the percentage of days in the test set when the true observation falls into the corresponding P.I. constructed for it. The empirical coverage rate of VaR is defined to be the percentage of days in the test set when the true observation is lower than the corresponding VaR constructed for it.} of the 95\% P.I. and 95\% VaR constructed by CuDvine and VAR, along with the corresponding $p$-values for the binomial test in Table \ref{SumCRPS}. {If the fitted model can approximate the multivariate time series well, the empirical coverage rates of both the constructed P.I. and VaR should be close to 95\%.} CuDvine gives an empirical coverage rate that is very close to the target rate~(95\%) and passes the binomial tests for both P.I. and VaR. \textcolor{black}{Neither VAR(1) or VAR(2) provides a satisfactory performance, while VAR(1)-DCC performs well for VaR but not for P.I.}

\begin{table}[]
	\centering
	\begin{tabu}{ccccc}\hline\hline	
		& CRPS    & QRPS   &VaR 95\% & P.I. 95\%         \\\hline
		CuDvine   & 0.161   & 0.040   & 93.15\%~(0.117) & 94.25\%~(0.471) \\
		VAR(1)      & 0.165   & 0.042  & 90.96\%~(0.001) & 89.86\%~(0)     \\
		\rowfont{\color{black}}
		VAR(2)      & 0.166   & 0.042  & 91.23\%~(0.002) & 89.04\%~(0)     \\
		\rowfont{\color{black}}
		VAR(1)-DCC      & 0.164   & 0.040  & 93.70\%~(0.278) & 90.41\%~(0)     \\\hline\hline
	\end{tabu}
	\caption{\it Average CRPS/QRPS for CuDvine and three VAR variants, and the empirical coverage rates of the one-day ahead 95\% VaR and 95\% P.I. for the demand-weighted price. The $p$-value of the corresponding binomial test is reported in the brackets.}
	\label{SumCRPS}
\end{table}

\textcolor{black}{In summary, the results in Table \ref{MarginalCRPS}-\ref{SumCRPS} clearly indicate that CuDvine has an edge over the VAR models in terms of prediction accuracy. Moreover, note that CuDvine is a parsimonious model with less parameters than the three VAR models. A few more observations can be drawn from the prediction results. First, the performance of VAR(1) is very similar to VAR(2), indicating its unfavorable performance is not caused by time lags. Second, VAR(1)-DCC performs the best among the three VAR models, showing evidence of time-varying dependence among the five regions. Third, despite the DCC specification, VAR(1)-DCC is still inferior to CuDvine by a wide margin, indicating the performance gain from CuDvine is not solely due to the time-varying cross-sectional dependence.}

\textcolor{black}{To further demonstrate the advantage of CuDvine, we compare the in-sample goodness of fit by CuDvine with the three VAR variants. Specifically, based on each estimated model, we simulate a time series $\{\bfY_t^{boot}=(Y_{t1}^{boot},\cdots,Y_{t5}^{boot})\}_{t=1}^{10000}$ of length 10000 and use it to numerically approximate the stationary distribution implied by the estimated time series model~(see Section \S5.2~(Scenario B) of the supplementary material for the detailed simulation algorithm). Two aspects of the multivariate time series are considered. First, for each univariate time series $\{Y_{ti}\}, i=1,\cdots,5$, we estimate the bivariate copula of its self-lagged pair $(Y_{ti}, Y_{t-1,i})$ based on the bootstrapped sample $\{\bfY_t^{boot}\}_{t=1}^{10000}$ via R package \texttt{kdecopula}, which provides kernel smoothing estimation for bivariate copula density. Second, using the same technique, we estimate the bivariate copula of cross-lagged pair $(Y_{ti}, Y_{t-1,j}), i\neq j$ based on the bootstrapped sample. For ground truth, we estimate the empirical bivariate copulas based on the observed multivariate time series of the training data $\{\bfY_t\}_{t=1}^{1460}$.}

\textcolor{black}{Figure \ref{NSW_VIC_Copula}(a)-(b) gives the contour plot of the estimated bivariate copula density (with standard normal margins) of self-lagged pair for NSW and VIC~(the result for other regions is similar and thus is omitted). As can be seen clearly, for both NSW and VIC, the copula implied by CuDvine best resembles the empirical copula. This is also confirmed by the corresponding Kendall's tau and Spearman's rho (provided on the plot) of each copula, where CuDvine provides the closest match to the empirical copula. Note that the empirical copulas assume irregular shapes and exhibit certain level of tail dependence, which explains the selection of non-Gaussian copulas~($t$- and Gumbel copula) by uDvines. Figure \ref{NSW_VIC_Copula}(c) gives the contour plot of the estimated copula of the cross-lagged pair (NSW, VIC), which again confirms the favorable performance of CuDvine.}

\section{Conclusion}
In this paper, we proposed and studied CuDvine -- a novel multivariate time series model that enables the simultaneous copula-based modeling of temporal and cross-sectional dependence for multivariate time series. We first studied a univariate time series model -- uDvine, that extends the first-order copula-based Markov chain to Markov chains of an arbitrary-order. By pair copula construction, uDvine provides flexible specifications for the marginal behavior and temporal dependence of univariate time series. To generalize to the multivariate context, we designed CuDvine by linking multiple uDvines via a copula. Compared to existing multivariate time series models, CuDvine shows greater balance between tractability and flexibility. We studied the probabilistic properties of uDvine in detail. We proposed a sequential model selection procedure and a two-stage MLE for the inference and estimation of CuDvine. The consistency and asymptotic normality of the MLE were formally established and affirmed by extensive numerical experiments. Finally, using applications on the Australian electricity price and the Ireland wind speed~(in the supplementary material), we demonstrated CuDvine's promising ability for modeling time-varying and spatio-temporal dependence of multivariate time series, and we observed significant improvement over traditional time series models in terms of prediction accuracy.


\clearpage

\bibliographystyle{apalike}
\bibliography{Reference}
\end{document}